\documentclass{pacificfusion}

\usepackage[numbers]{natbib}
\usepackage{xcolor}
\usepackage{appendix}
\usepackage{subcaption}
\usepackage{mwe}
\usepackage{graphicx}
\usepackage{xr}
\usepackage{marvosym}
\usepackage{xspace} 
\usepackage{booktabs} 

\newcommand{\micron}{\ensuremath{\mu\mathrm{m}}\xspace}

\title{Validation of FLASH for magnetically driven inertial confinement fusion target design}

\author[1]{C. Leland Ellison}
\author[2]{Jonathan Carroll-Nellenback}
\author[1]{Chiatai Chen}
\author[1]{Scott Davidson}
\author[1]{Bryan Ferguson}
\author[1]{Fernando Garcia-Rubio}
\author[2]{Edward C. Hansen}
\author[3]{Yannick de Jong}
\author[4]{Jacob R King}
\author[1]{Patrick Knapp}
\author[1]{Keith LeChien}
\author[1]{Anthony Link}
\author[1]{Nathan B. Meezan}
\author[1]{Douglas S. Miller}
\author[5]{Philip Mocz}
\author[2]{Kassie Moczulski}
\author[1]{Nantas Nardelli}
\author[1,2]{Adam Reyes}
\author[1]{Paul F. Schmit}
\author[1]{Hardeep Sullan}
\author[2]{Petros Tzeferacos}
\author[3]{Daan van Vugt}
\author[1]{Alex B. Zylstra}

\affil[1]{Pacific Fusion, 6082 Stewart Ave, Fremont, CA 94538}
\affil[2]{Flash Center for Computational Science, Department of Physics and Astronomy, University of Rochester, Rochester, NY 14611}
\affil[3]{Ignition Computing, Torenallee 30-12, 5617BD Eindhoven, NL}
\affil[4]{Fiat Lux, Lafayette, CO 80026 }
\affil[5]{Center for Computational Astrophysics, Flatiron Institute, 162 5th Ave, New York, NY 10010, USA}

\begin{abstract}
FLASH is a widely available radiation magnetohydrodynamics code used for astrophysics, laboratory plasma science, high energy density physics, and inertial confinement fusion. Increasing interest in magnetically driven inertial confinement fusion (ICF), including Pacific Fusion's development of a 60 MA Demonstration System designed to achieve facility gain, motivates the improvement and validation of FLASH for modeling magnetically driven ICF concepts, such as MagLIF, at ignition scale. Here we present a collection of six validation benchmarks from experiments at the Z Pulsed Power Facility and theoretical and simulation studies of scaling MagLIF to high currents. The benchmarks range in complexity from focused experiments of linear hydrodynamic instabilities to fully integrated MagLIF fusion experiments. With the latest addition of physics capabilities, FLASH now obtains good agreement with the experimental data, theoretical results, and leading ICF target design simulation code results across all six benchmarks. These results establish confidence in FLASH as a useful tool for designing magnetically driven ICF targets on facilities like Z and Pacific Fusion's upcoming Demonstration System.
\end{abstract}

\begin{document}
\maketitle

\section{Introduction}
\label{sec:introduction}
Multidimensional radiation hydrodynamics (rad-hydro) codes are an essential tool for designing ICF targets, planning experiments, and interpreting experimental data. Such codes have enabled progress on today's leading ICF facilities; for instance, the HYDRA multiphysics code was instrumental to obtaining ignition at NIF \cite{Marinak_2024, Kritcher_2024}. Similarly, pulsed magnetic fusion experiments at Z are regularly planned with high predictive capability using codes like HYDRA (e.g. Ref.~\cite{Ruiz_2023_Iscaling}), ALEGRA (e.g. Ref.~\cite{Knapp_2017}), and LASNEX (e.g. Ref.~\cite{Sinars_2011}). 

To date, the codes that have been most extensively validated and employed for ICF target design are non-public codes developed at national laboratories. Private-sector investments are now seeking to commercialize inertial fusion energy, creating a need for a widely-available and scientifically validated ICF target design code \cite{Ellison_2024}. This is occurring both for laser-based ICF concepts as well as magnetically driven ICF concepts, also known as ``current-driven ICF", ``pulsed magnetic fusion", ``magneto-inertial confinement fusion", or ``pulser ICF" in reference to pulsed-power drivers \cite{AMPS_2025}. In particular motivation for this work, Pacific Fusion requires a validated multi-physics code for designing magnetically driven ICF targets to field at its upcoming 60 MA Demonstration System \cite{AMPS_2025}. To this end, Pacific Fusion together with the Flash Center at the University of Rochester, Ignition Computing, Fiat Lux, and other collaborators have been developing, verifying, and validating the FLASH code for use as a ICF target design code for pulsed-power drivers.

FLASH \cite{Fryxell_2000} is a publicly available, massively parallel, multi-physics \cite{Tzeferacos_2015}, adaptive mesh refinement (AMR), finite-volume Eulerian hydrodynamics and magneto-hydrodynamics (MHD) code. It has a world-wide user base of more than 4,900 scientists, and more than 1,300 papers have been published using the code to model problems in a wide range of disciplines, including plasma astrophysics \cite{Tzeferacos_2018, Bott_2021}, supernova deflagrations \cite{Townsley_2007}, fluid dynamics, high energy density physics (HEDP) \cite{Fatenejad_2013, Sauppe_2023}, and Z-pinch plasmas \cite{Hansen_2024, Michta_2024, Tranchant_2025}. Over the past decade and under the auspices of the U.S. DOE NNSA, specific capabilities for ICF and high-energy-density physics have been added to FLASH \cite{Tzeferacos_2015}. Recent improvements include multiple state-of-the art hydrodynamic and MHD shock-capturing solvers \cite{Lee_2013}, three-temperature extensions \cite{Tzeferacos_2012} with anisotropic thermal conduction and high-fidelity magnetized heat transport coefficients \cite{Ji_2013}, electron-ion thermal equilibration, multi-group radiation diffusion, tabulated and multi-species EOS and opacities, extended MHD terms including the Nernst effect, laser energy deposition, circuit models, and numerous synthetic diagnostics \cite{Tzeferacos_2017}. 

These new and existing features provided an excellent foundation for building a magnetically driven ICF target simulation capability, but additional code development and validation was needed before gaining confidence that FLASH was able to accurately simulate magnetically driven fusion targets of interest. Additional capabilities developed in support of this work include hydrogenic thermonuclear burn, a preheat energy source, alpha particle energy diffusion and deposition, multi-species models for material properties like electrical and thermal conductivity, and improved treatments of magnetic diffusion especially in the presence of vacuum/conductor interfaces and AMR refinement boundaries. A subset of these capabilities has been contributed to the development version of FLASH to be made available for future release (exceptions include the hydrogenic burn and alpha diffusion packages). Explanations of the system of partial differential equations solved by FLASH together with details of its constitutive (material) models and numerical methods are provided in Section \ref{sec:background}. Individual physics capabilities were verified using focused-physics test problems, often with exact analytic solutions for comparisons. Many such problems are described in the FLASH User Guide \cite{FLASH_UserGuide_2024}, including iconic MHD benchmarks such as Brio-Wu, Orszag-Tang, and MHD Rotor. In the course of this work, we supplemented the Flash Center's verification problems with additional problems, including solutions to diffusive processes in cylindrical geometries in the presence of sharp material interfaces such as occur at vacuum/conductor boundaries. The verification problems are included in automated testing pipelines to ensure that correctness is maintained over time. 

For validation of FLASH for magnetically driven ICF targets, we identified a collection of six benchmarks involving a combination of experimental data, simulation results from state-of-the-art ICF target design codes, and theoretical studies published in the peer-reviewed scientific literature. These benchmarks are:
\begin{enumerate}
\item Magneto-Rayleigh Taylor Growth Platform on Z~\cite{Sinars_2010, Sinars_2011}
\item Unmagnetized MagLIF Be Liner Platform on Z~\cite{McBride_2012, McBride_2013}
\item Richtmyer-Meshkov Growth Platform on Z~\cite{Knapp_2020}
\item ICF Confinement Time Platform on Z~\cite{Knapp_2017}
\item MagLIF Z Shot 2977 (e.g., Ref.~\cite{Lewis_2023})
\item MagLIF Scaling Study to High Current~\cite{Ruiz_2023_Iscaling}
\end{enumerate}
These benchmarks were chosen to span focused-physics experiments, such as key hydrodynamic instability dynamics, as well as high-performing fusion platforms including MagLIF at Z scale and beyond. With the exception of Benchmark 6, these benchmarks all include experimental data from targets fielded at the Z facility at Sandia National Laboratories. Diagnostic data used for our simulation comparisons include photon Doppler velocimetry (PDV) probes for liner position and acceleration, x-ray radiography, and nuclear burn diagnostics. Most of the benchmarks also include published simulation results from the target design codes that were used in pre- and/or post-shot modeling of these experiments, allowing comparison of the FLASH simulation results to the design codes used to field these experiments. The scaling study in Benchmark 6 includes both theoretical scaling and multi-dimensional HYDRA simulation results allowing us to assess FLASH’s validity in high current regimes relative to the best present scientific understanding (i.e., comparing against the code used to design igniting targets at NIF).

The following sections review the capabilities of the FLASH code (Section~\ref{sec:background}) and define and present the validation benchmarks (Section~\ref{sec:validation}). In all cases, we find good agreement between FLASH and the benchmark data from experiments, simulations, and theory. This includes matching the yield and burn-averaged ion temperature of MagLIF shot 2977 on Z and good agreement with HYDRA when modeling similarity scaled MagLIF targets from 15 MA to 60 MA in both alpha-deposition-off and alpha-deposition-on scenarios including targets exhibiting ignition and significant alpha heating. In aggregate, this validation study establishes confidence that FLASH may be used to design magnetically driven ICF targets at Z scale and ignition scale including that of Pacific Fusion's upcoming Demonstration System.

\section{Background}
\label{sec:background}

FLASH is a multi-physics, adaptive mesh refinement, finite-volume Eulerian, multi-group radiation magnetohydrodynamics code. It solves the hydrodynamic description of a magnetized plasma including a three-temperature extension for electron, ion and radiation temperature fields. It can handle multiple ion species assumed to be in local thermodynamic equilibrium.

\subsection{Governing Equations}
\label{ssec:governing}
The relevant system of equations that FLASH solves corresponds to continuity, momentum and total energy conservation, together with magnetic induction, reading:

\begin{equation}
    \frac{\partial \rho}{\partial t} + \nabla\cdot\left(\rho\boldsymbol{v}\right) = 0,
    \label{eq:continuity}
\end{equation}

\begin{equation}
    \frac{\partial}{\partial t}\left(\rho \boldsymbol{v}\right) + \nabla\cdot\left(\rho\boldsymbol{v}\boldsymbol{v} + p_{\text{tot}}\overline{\overline{\mathbf{I}}} - \boldsymbol{B}\boldsymbol{B}\right) = 0,
    \label{eq:momentum}
\end{equation}

\begin{equation}
     \frac{\partial}{\partial t}\left(\rho E_{\text{tot}}\right) + \nabla\cdot\left[\left(\rho E_{\text{tot}} + p_{\text{tot}} \right)\boldsymbol{v} - \boldsymbol{B}\left(\boldsymbol{v}\cdot\boldsymbol{B}\right)\right] = - \nabla\cdot\boldsymbol{q}  + \nabla\cdot\left[\boldsymbol{B}\times\left(\eta \nabla\times\boldsymbol{B}\right)\right] + Q_{\text{las}},
     \label{eq:energy}
\end{equation}

\begin{equation}
    \frac{\partial \boldsymbol{B}}{\partial t} + \nabla\cdot\left( \boldsymbol{v B} - \boldsymbol{Bv}\right) = -\nabla\times\left(\eta\nabla\times\boldsymbol{B}\right).
    \label{eq:induction}
\end{equation}

Here, $\rho$ is the total mass density, $\boldsymbol{v}$ is the average fluid velocity, $\boldsymbol{B}$ is the magnetic field normalized with the factor of $\sqrt{4\pi}$, 
 
\begin{equation}
    p_{\text{tot}} = p_{\text{ele}} + p_{\text{ion}} + p_{\text{rad}} + \dfrac{B^2}{2},
\end{equation}

\begin{equation}
    E_{\text{tot}} = \dfrac{1}{2}v^2 + e_{\text{ele}} + e_{\text{ion}} + e_{\text{rad}} + \dfrac{1}{2}\dfrac{B^2}{\rho},
\end{equation}

are the total pressure and specific total energy, which includes the specific internal energies of the electrons, ions, and radiation field along with the specific kinetic and magnetic energies. Here, $B$ and $v$  are the magnitudes of the vector quantities in bold font $\boldsymbol{B}$ and $\boldsymbol{v}$,  respectively. The total heat flux accounts for the electron and ion conductivities as well as the radiation energy flux, $\boldsymbol{q} = \boldsymbol{q}_{\text{ele}} + \boldsymbol{q}_{\text{ion}} + \boldsymbol{q}_{\text{rad}}$, $\eta$ refers to the magnetic resistivity, and $Q_{\text{las}}$ represents a volumetric external energy deposition (laser). 

When adopting a three-temperature model, the above system is complemented with the following three equations, corresponding to the change in specific energies of the electrons, ions, and radiation field:

\begin{equation}
    \frac{\partial}{\partial t}\left(\rho e_{\text{ele}}\right) + \nabla\cdot\left(\rho e_{\text{ele}}\boldsymbol{v}\right) + p_{\text{ele}}\nabla\cdot\boldsymbol{v} = -\nabla\cdot\boldsymbol{q}_{\text{ele}} +\rho\dfrac{c_{v,\text{ele}}}{\tau_{ei}}\left(T_{\text{ion}} - T_{\text{ele}}\right) +  Q_{\text{abs}} - Q_{\text{emis}} + Q_{\text{ohm}} + Q_{\text{las}},
    \label{eq:energy_ele}
\end{equation}

\begin{equation}
    \frac{\partial}{\partial t}\left(\rho e_{\text{ion}}\right) + \nabla\cdot\left(\rho e_{\text{ion}}\boldsymbol{v}\right) + p_{\text{ion}}\nabla\cdot\boldsymbol{v} = -\nabla\cdot\boldsymbol{q}_{\text{ion}} +\rho\dfrac{c_{v,\text{ele}}}{\tau_{ei}}\left(T_{\text{ele}} - T_{\text{ion}}\right),
    \label{eq:energy_ion}
\end{equation}

\begin{equation}
    \frac{\partial}{\partial t}\left(\rho e_{\text{rad}}\right) + \nabla\cdot\left(\rho e_{\text{rad}}\boldsymbol{v}\right) + p_{\text{rad}}\nabla\cdot\boldsymbol{v} = -\nabla\cdot\boldsymbol{q}_{\text{rad}} -  Q_{\text{abs}} + Q_{\text{emis}}.
    \label{eq:energy_rad}
\end{equation}
 
In these equations,  $c_{v,\text{ele}}$ is the electron specific heat, $\tau_{ei}$ is the ion/electron equilibration time, $Q_{\text{abs}}$ and $Q_{\text{emis}}$ are the increase/decrease in electron internal energy due to the total absorption/emission of radiation, while $Q_{\text{ohm}}$ represents the Ohmic heating of the plasma (equal to the second term in the right-hand side of the total energy equation \eqref{eq:energy}). In the continuum, these three equations are redundant with the total energy equation. For a discussion of how FLASH updates this system, see \cite{Tzeferacos_2015} or the publicly available FLASH Users' Guide, Subsection 14.1.5.2: the RAGE-like Approach \cite{FLASH_UserGuide_2024}.

The three-temperature equations of state in FLASH relate the internal energies, pressures, and temperatures of the components. It is implicitly assumed that species are in local thermodynamic equilibrium. Species pressures are added via the partial pressures approximation with each species EOS evaluated at its partial density. We use tabular EOS data from common databases such as SESAME and PROPACEOS in our benchmark problems. FLASH has the capability to advect mass scalars, including species mass fractions. Although FLASH can model microscopic mixing using a diffusion velocity given by Fick’s law, we are currently not utilizing this capability due to the absence of a mass diffusion coefficient implementation for multi-species plasmas. Instead, species are subjected to numerical diffusion operating at the grid scale which has been minimized by selecting a fine enough grid resolution that quantities of interest are converged.

\subsubsection{Diffusive processes}
\label{ssec:diffusive}

FLASH uses diffusion schemes, moderated
by appropriate flux limiters, to approximate several physical transport processes; ion and electron heat conduction, magnetic diffusion, radiation transport, and alpha energy transport. The simulation’s evolution – driven by the terms on the right-hand side of the governing equations – is carried out under the assumption of constant density during the operator-split diffusive process. 

Ions and electrons exchange internal energy through collisions at a rate given by a Spitzer ion-electron equilibration timescale but with the Coulomb logarithm from Lee $\&$ More expressions \cite{Lee_1984}. The electron and ion heat fluxes are proportional to the gradient of temperature through their respective conductivities, 
\begin{equation}
    \boldsymbol{q}_{\text{ele,ion}} = - K_{\text{ele,ion}}\nabla T_{\text{ele,ion}}.
    \label{eq:qeleion}
\end{equation}
We incorporate a Bohm-like scaling to account for magnetization effects, which replaces the familiar Braginskii magnetized transport scaling for strongly magnetized plasmas with a $1/\left(\omega \tau\right)$ scaling based on the addition of Bohm collisions $\nu _{\text{bohm} = 16\omega_{\text{ce}}}$. This is often used in MagLIF simulations in the literature \cite{slutz2010pulsed}. For ion conductivity, we rely on analytical Braginskii expressions. 

We next describe the contribution due to radiation transport. FLASH incorporates radiation effects using multigroup diffusion (MGD) theory. The total specific radiation energy in Eq. \eqref{eq:energy_rad} is divided into contributions from the radiation energy density of each group $u_g$, 

\begin{equation}
    \rho e_{\text{rad}}=\sum _{g=1}^{N_g}u_g.
\end{equation}
Similarly, the total radiation flux, emission, and absorption terms that appear in the radiation energy equation \eqref{eq:energy_rad} contain contributions from each energy group. The change is the radiation energy density for each group is described by 

\begin{equation}
    \frac{\partial u_g}{\partial t} + \nabla\cdot\left(u_g\boldsymbol{v}\right) + \left(\dfrac{u_g}{\rho e_{\text{rad}}}\right)p_{\text{rad}}\nabla\cdot\boldsymbol{v} = -\nabla\cdot\boldsymbol{q}_{g} -  Q_{\text{abs},g} + Q_{\text{emis},g}, 
\end{equation}
where the radiation flux, absorption and emission per group are given by

\begin{equation}
    \boldsymbol{q}_{g} = - \dfrac{c}{3\kappa _{r,g}}\nabla u_g,\quad Q_{\text{abs},g} = c\kappa _{p,g} u_g, \quad Q_{\text{emis},g} = 4\kappa_{p,g}\sigma T_{\text{ele}}^4\dfrac{15}{\pi^4}\left[P\left(x_{g+1}\right)-P\left(x_{g}\right)\right].
\end{equation}

Here, $c$ and $\sigma$  are the speed of light and the Stephan-Boltzmann constant, respectively, and $P(x)$  is the Planck integral of the normalized photon energy $x=h\nu/k_BT_{\text{ele}}$. We have calculated the  Rosseland and Planck opacities per group, $k_{r,g}$  and $k_{p,g}$, using publicly available tabular TOPS opacities \cite{Magee_1995}. Flux limiters can be used to restrict the radiation flux to the free-streaming limit of photons. Benchmarks 1-4 introduced in Sec. \ref{sec:introduction} use gray (single group) diffusion while Benchmarks 5 and 6 were run with multigroup configurations.

Pacific Fusion developed internal burn capabilities absent in the public FLASH version, in which a diffusion equation for alpha particle energy is advanced in time self-consistently with the rest of the governing equations. We employ the Bosch $\&$ Hale formulas \cite{BoschHale1992} for the fusion reactivity. Deposition on the background fluid elements is given by an analytic rate and a user-defined partitioning between the electron and ion energies. 

All of the diffusive processes discussed – heat conduction, magnetic diffusion, radiation transport, and alpha energy diffusion – are solved implicitly using a backward-Euler scheme to avoid time-step constraints in highly diffuse regions. FLASH employs the HYPRE linear algebra package~\cite{Falgout_2006} to solve the system of equations that arises from discretizing each diffusion equation in zone-centered grid points on the AMR grid. HYPRE makes many standard iterative linear solver algorithms available, including PCG and GMRES together with preconditioners such as AMG.

\subsubsection{Extended MHD terms}

Extended MHD effects, especially Nernst, can be significant in the parameter regimes relevant to MagLIF \cite{Slutz_PoP_2010}. We have included Nernst advection of the magnetic field in Benchmarks 5 and 6. We use Braginskii expressions for the thermoelectric coefficient \cite{Braginskii1965}, which accounts for the reduction in the Nernst velocity due to magnetization. We found it beneficial to upwind the Nernst velocity computation to mitigate numerical instabilities occasionally observed when compressional waves converge on axis in a magnetized plasma, leading to the creation of spots with negative axial magnetic field values. 

\subsubsection{Energy deposition source}

Although FLASH offers a laser ray-tracing package \cite{FLASH_UserGuide_2024}, a simplifying approach to modeling MagLIF experiments is to deposit energy at a fixed rate across a given volume or mass. For the MagLIF benchmarks in this work, the laser deposition term, $Q_{\text{las}}$ in Eq. \eqref{eq:energy}, is modeled as a prescribed external source that deposits energy at a constant rate, either per unit volume or mass, within the region defined by the preheat radius and during the preheat time. This is also the approach used in the reference publication for the MagLIF scaling study in Benchmark 6 \cite{Ruiz_2023_Iscaling}.

\subsubsection{Vacuum settings}

A high magnetic resistivity is applied in the vacuum region to ensure it remains current-free, allowing the magnetic field to diffuse rapidly from the boundary condition to the liner's outer surface. Grid cells with density values below some user-specified value are assigned a large magnetic resistivity value to rapidly diffuse magnetic field and carry minimal current. 

\subsection{Discretization and time evolution}
\label{ssec:discretization}

FLASH discretizes the equations in Subsection~\ref{ssec:governing} onto a Cartesian, cylindrical or spherical orthogonal grid. Computational cells are conceptually organized into blocks, with typical FLASH simulations containing eight inner cells per block along each linear dimension. FLASH employs the PARAMESH package \cite{Olson_1999, MacNeice_2000} to implement the adaptive mesh refinement (AMR) grid. The scheme employed is the block-structured oct-tree AMR, sharing features with other approaches in the literature (see, e.g., \cite{BergerOliger1984, BergerColella1989, Khokhlov1998}). The fundamental data structure is a block of cells, arranged in a logically Cartesian fashion, which implies that each cell can be specified using a unique block identifier and, e.g., a coordinate doublet (viz., i and j) or triplet (viz., i, j, and k) in two and three dimensions, respectively. This indexing scheme does not require a physically rectangular coordinate system and can be applied directly to curvilinear coordinates. The complete computational grid consists of a collection of blocks with different physical cell sizes, which are related to each other in a hierarchical fashion using a tree data structure. The blocks at the root of the tree have the largest cells, while their ``child'' blocks have smaller cells and are said to be refined. The refinement of a region in space is triggered by a refinement criterion, which in PARAMESH is adapted from L\"ohner’s \cite{Lohner_1987} error estimator. The latter is a dimensionless quantity that can be applied with complete generality to any of the field variables of the simulation or any combination thereof, based on a discrete, modified second derivative, normalized by the average of the gradient over one computational cell. In one dimension, this would read: 
\begin{equation}
E_i = \frac{|u_{i+2}-2u_{i}+u_{i-2}|}{|u_{i+2}-u_{i}| + |u_i-u_{i-2}|+ \epsilon[|u+{i+2}|+2|u_i|+|u_{i-2}|]}
\end{equation}
where $u_i$ is the refinement test variable value in the i-th cell. The last term in the denominator of this expression acts as a filter that prevents the refinement of small ripples, where $\epsilon$ is a small constant. We generally choose density and fuel concentration as refinement variables. We minimize the refinement level encountered in the vacuum region by only enabling higher refinement levels in the fuel.

A series of operator splits is used to evolve the governing equations in time. All the terms on the left-hand side describe the advection of conserved quantities and the effect of work. FLASH advances them in time through the use of a Godunov scheme \cite{Godunov_1959} based on PPM \cite{Colella_1984} with CTU for high-order Riemann states \cite{Colella_1990, Lee_2009} and approximate Riemann solvers for upwinded fluxes. Our benchmarks are typically run without additional artificial viscosity. In the two-dimensional runs described, in-plane magnetic fields are evolved with the constrained transport method on a staggered mesh to preserve their divergence-free character during the hydro update \cite{Evans_1988, Lee_2009}. 

\subsubsection{Magnetic boundary conditions}

FLASH offers two methods of coupling magnetic energy into the simulation domain, both of which manifest as a boundary condition for the azimuthal magnetic field $B_\theta$ at the outer radial boundary. All of the benchmark simulations take place in 1D $(r)$ and 2D $(r,z)$ cylindrical geometry. The assumption of azimuthal symmetry allows us to use Amp\`ere’s Law to relate the magnetic field at the outer boundary $r_0$ to the total amount of current $I$ enclosed within the Amperian loop of radius $r_0$ at that location: $B_\theta = 2I/r_0 c$. This equation is used in both of the circuit coupling methods available in FLASH.

In the first method, the user defines a current as a function of time for the simulation duration: $I(t)$. Most commonly, this is given in a tabular file format as a series of time-current pairs. The boundary condition at a given time t is then computed according to the Amp\`ere's Law boundary condition above by linearly interpolating the table to the desired instant in time.

This method is most suitable when comparing against an experiment with an experimentally measured current profile. We use it in Benchmarks 1-4. It has the drawback that the current drive is unaware of any impedance change in the simulation domain; in effect, the circuit is able to provide arbitrarily large voltage to ensure the specified current is present in the simulation domain. When it is important to incorporate the effects of the dynamic impedance, the second method is preferable.

In the second method, the MHD simulation domain is coupled to an external set of ordinary differential equations that model the evolution of a lumped-element circuit model. This is used in Benchmarks 5 and 6. A system of ODE's is constructed for the network of circuit elements using Kirchhoff’s laws. Examples of systems of circuit ODE's in FLASH are present in the FLASH User Guide \cite{FLASH_UserGuide_2024}. The equations are differenced using a backward Euler scheme. In all cases, the voltage across the load $V_{\text{load}}$ appears as an unknown in the circuit equations, where the ``load" is represented by the MHD simulation domain. In the differenced equations, given $V_{\text{load}}$ at a particular instant in time, one can solve for the circuit state at the new time $t + \text{d}t$, including the current through the load I. This current can then be used as the boundary condition for the MHD simulation for the next timestep. To compute the impedance of the simulation domain $V_{\text{load}}$, FLASH uses Faraday’s Law by tallying the change in magnetic flux in the $r,z$ plane over a given time interval.  This circuit coupling model was validated by performing high accuracy integrations of the circuit ODE assuming limiting cases of constant inductance and inductance varying in time in some known manner, then comparing to FLASH simulations where hydrodynamic motion was frozen or prescribed to give a load inductance of the desired form. Good agreement was obtained. 

\section{Validation Benchmarks}
\label{sec:validation}

\subsection{Benchmark 1: Single-Mode Magneto-Rayleigh-Taylor}
\label{ssec:single-mode_mrt}

An important consideration for magnetically driven ICF targets is robustness with respect to the Magneto-Rayleigh-Taylor (MRT) instability \cite{Kruskal_1954, Harris_1962}. In effect, the magnetic acceleration manifests an RT-unstable interface at the outer surface of the liner; the low mass density of the vacuum ``fluid" results in a large Atwood number for the instability. Any perturbations of this interface along the axial dimension will therefore grow exponentially in time (in the linear regime). When designing high performing fusion targets, the liner must be sufficiently thick to prevent the MRT perturbations from causing non-uniform compression of the fuel or breakup of the liner.

\begin{figure}
    \centering
    \includegraphics[width=0.8\linewidth]{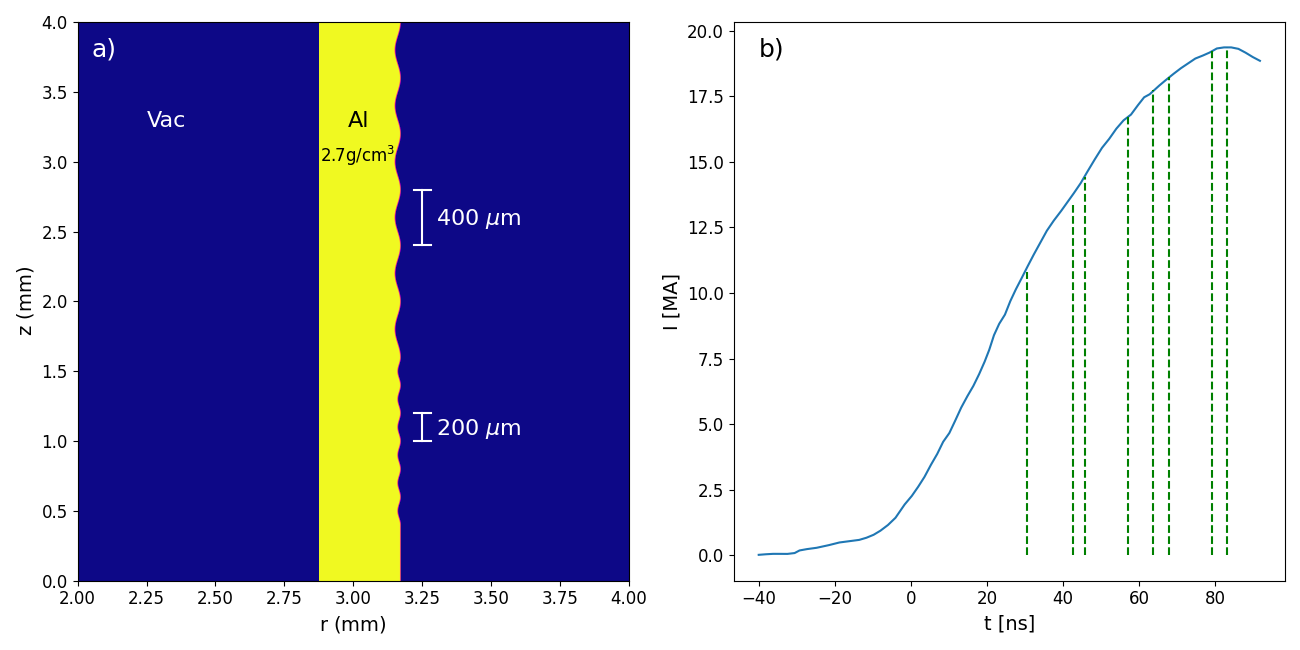}
    \caption{(a) Schematic of the aluminum liner with pre-imposed sinusoidal perturbations used for the MRT benchmark experiments at Z. (b) Current trace from the aluminum liner (Series 1)  MRT experiment, digitized from Ref.~\cite{Sinars_2011}; the times of the radiographs are shown with the vertical lines.}    
    \label{fig:sinars_mrt_recap}
\end{figure}

To provide experimental benchmark data for this instability in the context of magnetically driven ICF, a series of experiments were conducted at Z wherein metallic liners were machined with sinusoidal perturbations to seed MRT growth of readily diagnosable size \cite{Sinars_2010, Sinars_2011, Sinars_2012}. Beryllium and aluminum liners were fielded with perturbations of varying wavelengths and in-flight radiographs were used to diagnose the time evolution of the perturbation. A prototypical current trace and pre-shot photograph of an aluminum liner are shown in Fig.~\ref{fig:sinars_mrt_recap}. Over the years, these experiments have been simulated using several codes including LASNEX and Gorgon, which are commonly used to design experiments at Z. 

For this benchmark, we focus our attention on the Series 1 experiments, which involve an aluminum liner machined with two perturbations on different portions of the target. One portion of the target has a 400-\micron wavelength, 20-\micron amplitude and the other has a 200-\micron wavelength, 10-\micron amplitude perturbation. The Series 1 current drive shown in Fig.~\ref{fig:sinars_mrt_recap} and peaks at approximately 20 MA with a rise time of $\approx90$ ns. The inner radius of the target is 2.876 mm and the outer radius is 3.168 mm; the sinusoidal perturbation is imposed from this outer radius, i.e.
\begin{equation}
    r(z) = r_o + \delta \sin\left( \frac{2 \pi z}{\lambda}\right)
\end{equation}
where  $r_o$ is the outer radius given above,  $\delta$ is the amplitude, and $\lambda$ is the wavelength. To measure the evolution of the perturbation in the experiments, 6.151 keV x-ray radiographs were taken at 30.5, 42.7, 45.8, 57.0, 63.6, 67.7, 79.0, and 83.0 ns in reference to the time shown on the current trace on Fig.~\ref{fig:sinars_mrt_recap}. 

We conducted FLASH simulations of this platform using the multiphysics capabilities described in Section~\ref{sec:background}. For simplicity, we performed separate simulations of the 400-\micron and 200-\micron perturbation. In both cases we simulated 0.08 mm of axial length of the target with periodic boundary conditions and radially from r=0 to 6 mm. Periodic boundary conditions were applied in the axial direction for these “short” 2D target segments. We used five levels of adaptive mesh refinement to follow the density interfaces with a resolution of dx = dy = 3.125 \micron at the finest level. To supply magnetic energy to the problem we used a current-driven boundary condition, as explained in Sec.~\ref{sec:background}, with the current pulse shown in Fig.~\ref{fig:sinars_mrt_recap}. 

\begin{figure}
    \centering
    \includegraphics[width=0.8\linewidth]{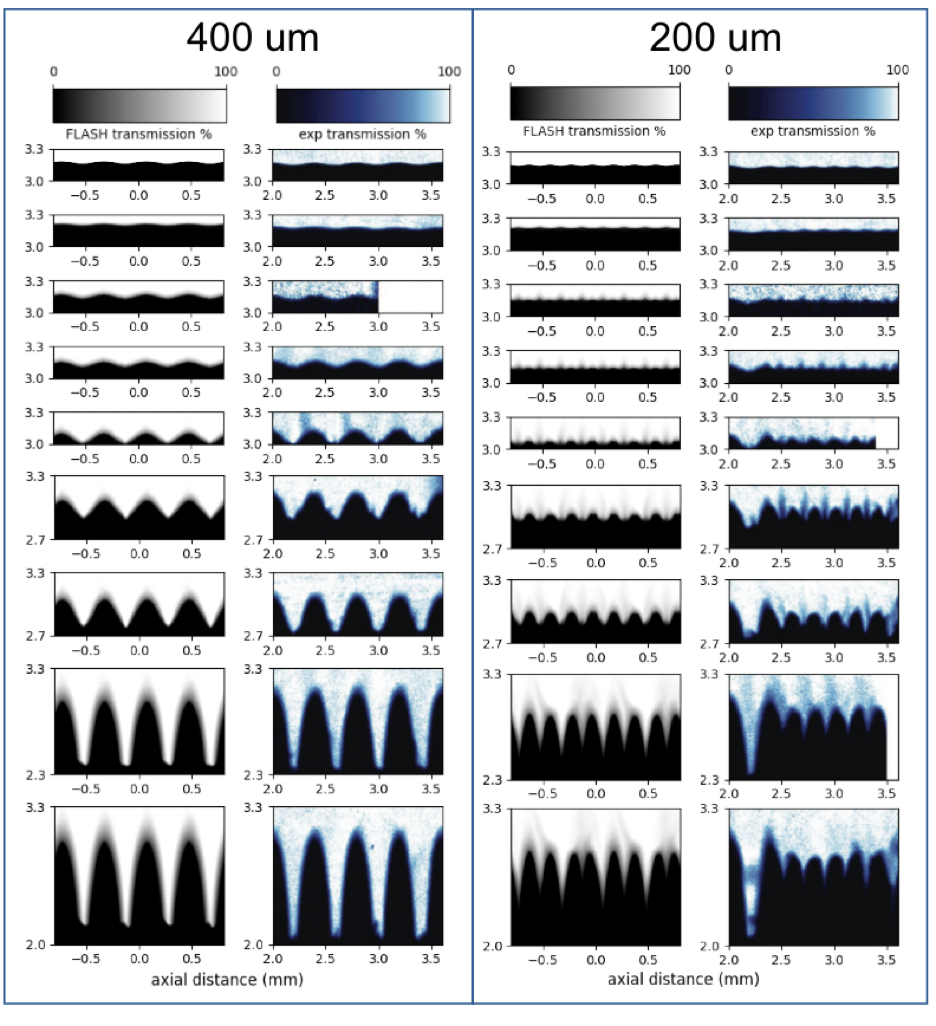}
    \caption{Experimental x-ray radiographs of the 400-\micron and 200-\micron wavelength perturbed aluminum liners for the MRT benchmark platform at Z \cite{Sinars_2011} compared to synthetic transmission ratios from FLASH simulations using the cold opacity for aluminum and a synthetic ray tracing method. Note that the 200-\micron wavelength experimental data includes a portion of the 400-$\mu$m wavelength segment of the target, so the deep groove structure emerging near ~2.2 mm in the experimental radiographs is an artifact of this. Radiographs and simulation times correspond to 30.5, 42.7, 45.8, 57.0, 63.6, 67.7, 79.0, and 83.0 ns going from top to bottom. See Fig.~\ref{fig:sinars_mrt_recap} for timings relative to the current pulse. Experimental radiographs from Fig.~7 of Ref.~\cite{Sinars_2011}, licensed under a Creative Commons Attribution (CC BY) license.}
    \label{fig:sinars_mrt_sim_radiographs}
\end{figure}

To analyze the results we first generated synthetic radiographs for comparison with the experimental data. This comparison is shown in Fig.~\ref{fig:sinars_mrt_sim_radiographs}. For the FLASH data, we used the mass density to compute a transmission percentage for 6.151 keV x-rays using a ray tracing algorithm and the cold opacity for aluminum; although the outer portions of the liner undergo heating and phase changes, the cold opacity can often be sufficient for comparisons against experimental data. In the Figure we truncate the spatial window to match the axial and radial extents of the experimental data at the corresponding time. We leveraged the periodic assumption in our simulation to extend the axial length to the full range. In the synthetic analysis the detector was simulated at a three-degree angle relative to the perpendicular direction from the axial direction of the target, which was chosen to match the alignment of the experimental detector. 

The FLASH data shows excellent agreement with the experiments in terms of implosion timing and mode amplitude growth.  Note also that the 200 $\mu$m wavelength experiment exhibits jetting features at the nodes of the sinusoidal perturbation. Similar jetting features are present in the FLASH simulations although the densities appear somewhat less than those experimentally observed. Importantly, no jets were observed in either the experiment or FLASH simulations at 400 \micron wavelengths.

\begin{figure}
    \centering
    \includegraphics[width=0.5\linewidth]{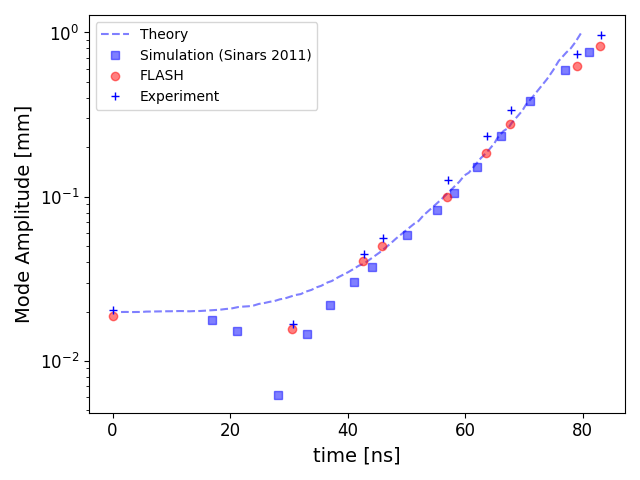}
    \caption{Mode amplitude of the 400-$\mu$m wavelength perturbation on the aluminum liner MRT benchmark experiments from Ref.~\cite{Sinars_2011}. The FLASH mode amplitude, extracted as the difference between the maximum and minimum radii of a 0.1 g/cm$^3$ density contour, agrees well with the experiment and theory. The level of agreement is consistent with those obtained by the code used in the referenced publication \cite{Sinars_2011}.}
    \label{fig:sinars_mrt_amplitude}
\end{figure}

To compare against theory, we extracted effective mode amplitudes from the times of the experimental radiographs from the FLASH simulations. The mode amplitude was quantified by constructing a contour of constant density $\rho = 0.1$ g/cm$^3$ and extracting the maximum and minimum radii at which that density occurred along the vacuum/conductor interface. The mode amplitude was then taken to be the difference between these maximum and minimum radii. The inferred mode amplitude as a function of time was not very sensitive to the exact choice of density threshold. The results of the comparison are shown in Fig.~\ref{fig:sinars_mrt_amplitude}. Also included are the linear growth rate and the rad-hydro simulations from Ref.~\cite{Sinars_2011}. 

Between the radiographs and the inferred growth rates, the FLASH data agrees well with the experimental, simulation, and theoretical results. Notably, the simulated mode amplitude is captured just as accurately as that in the published simulation data from the LASNEX code, commonly used to design targets for the Z facility. Accurately capturing the linear evolution of the MRT instability is a fundamental ingredient for modeling liner-based magnetically driven ICF targets. 

\subsection{Benchmark 2: Multi-mode Magneto-Rayleigh-Taylor}
\label{ssec:multi-mode_mrt}

In this benchmark we validate FLASH’s ability to capture representative MRT growth emerging from fine-scale material imperfections – surface roughness, voids, and inclusions – in metallic liner experiments. This test is more sophisticated than the previous one in that it involves the nonlinear dynamics of the hydrodynamic phenomenon and fine-scale spatial structures. It is also more relevant to magnetically driven fusion targets, which exhibit such multimode, nonlinear MRT behavior. In this benchmark, we compare FLASH with experimental radiographs and find that seeding the perturbation with a random temperature perturbation on the initial condition is able to recover good agreement with the experimental data.

Leading up to fully integrated MagLIF experiments, Sandia National Laboratories conducted hydrodynamic benchmark experiments with imploding beryllium liners without any laser preheat or pre-imposed axial magnetization \cite{McBride_2012, McBride_2013}. Beryllium liners with an initial inner radius of 2.8875 mm and outer radius of 3.4688 mm (i.e., aspect ratio $= 6$) were imploded with $\approx$20 MA, 100-ns rise time pulse and radiographic measurements were made to diagnose the in-flight MRT growth. A key motivation for these experiments was to assess the magnitude of MRT perturbations at the inner liner surface. The radiographic measurements provide time-resolved measurements of the multi-mode MRT structures that grow out of the target’s surface roughness and other manufacturing imperfections. 

\begin{figure}
    \centering
    \includegraphics[width=0.8\linewidth]{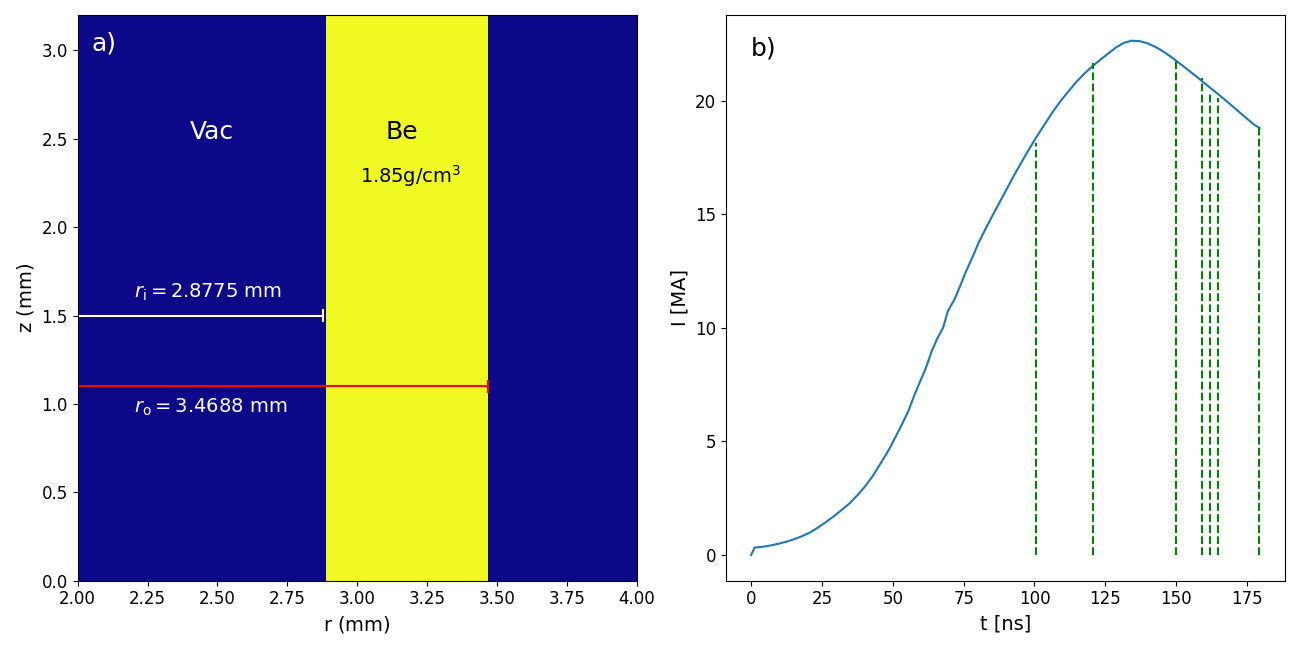}
    \caption{(a) Schematic showing the target configuration of the imploding beryllium liner experiments of Ref.~\cite{McBride_2013}. (b) Current trace z2173 with vertical bars indicating the time of the radiographs taken on those experiments.}
    \label{fig:mcbride_target_schematic}
\end{figure}

Figure~\ref{fig:mcbride_target_schematic} shows the experimental configuration and the drive current inferred from z2173 in Ref.~\cite{McBride_2012}. Across the six experiments, seven radiographs were taken at 100.4, 120.5, 149.8, 159.1, 162.1, 164.9 and 179.1 ns, which are depicted as vertical lines in the figure. 

We simulated a representative experiment using the current trace from shot z2173 (published in Ref.~\cite{McBride_2012}) in FLASH. To mimic the effects of surface roughness in seeding the multi-MRT structures, we applied a random temperature perturbation to the initial temperature of the liner of the form:
\begin{equation}
    T_\mathrm{ele}(r_\mathrm{i}, z_\mathrm{j}) = T_\mathrm{ion}(r_\mathrm{i}, z_\mathrm{j}) = \max \left(T_\mathrm{min}, T_0 + \exp^{(r_\mathrm{i} - r_\text{o})/\lambda_r} N_{\delta T}(i,j)\right)
    \label{eq:tele_perturbation}
\end{equation}
where $T_\mathrm{ele}(r_\mathrm{i}, z_\mathrm{j})$ is the electron temperature at the zone centered at radius $r_\mathrm{i}$ and axial position $z_\mathrm{j}$ and similarly for the ion temperature $T_\mathrm{ion}$. $T_0 = 293\ \text{K}$, $T_\textrm{min} = 273\ \text{K}$ is a temperature floor,  $\lambda_r$ is the perturbation decay length, $r_\mathrm{o} = 3.47$ mm is the outer radius of the liner and $N_{\delta T}(i,j)$ is a random number drawn from a normal distribution with mean 0 and standard deviation $\delta T$. 

\begin{figure}
    \centering
    \includegraphics[width=0.5\linewidth]{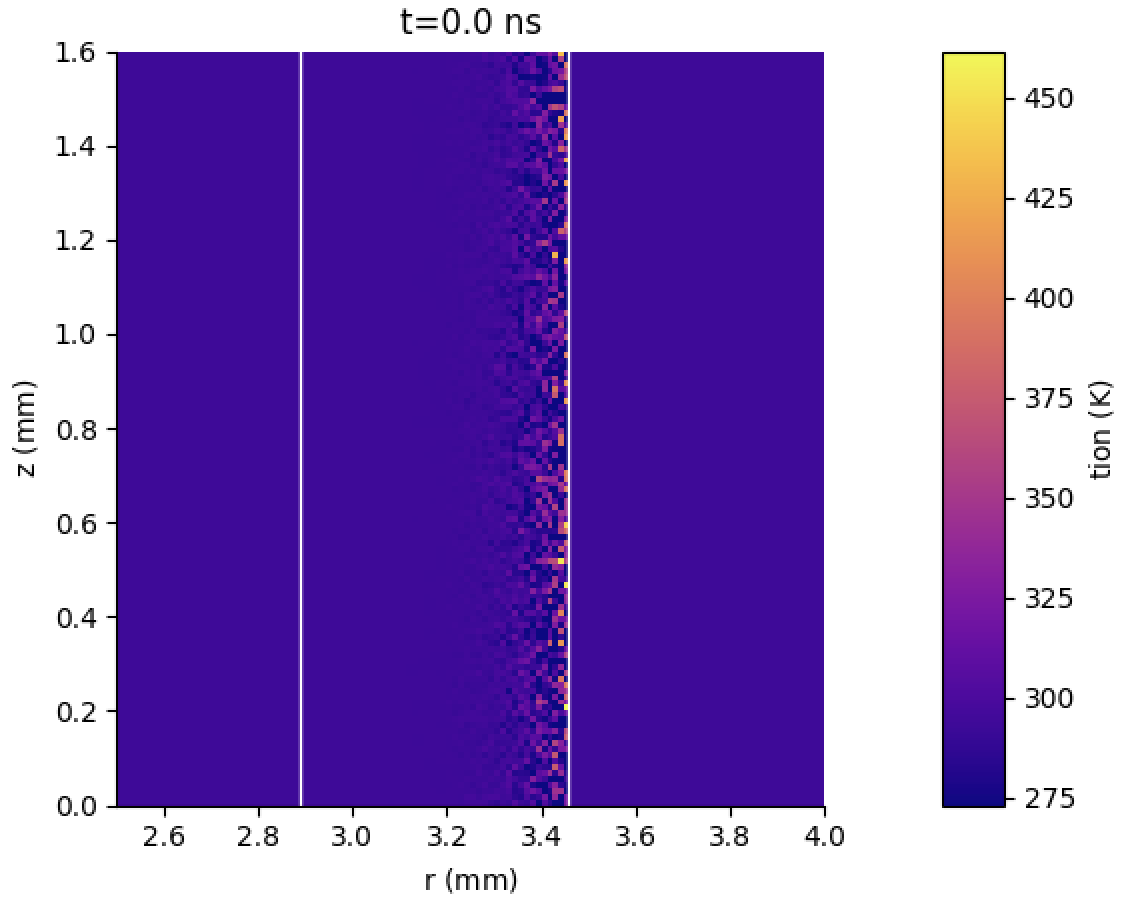}
    \caption{FLASH initial condition of the temperature in K for the multimode MRT benchmark according to Eq.~\eqref{eq:tele_perturbation}. This random temperture perturbation breaks the axial symmetry to seed ETI and MRT growth and is qualitatively representative of non-uniform early time heating due to defects in the target material (surface roughness, voids and inclusions). See, e.g., Figure 82 of Ref.~\cite{Hutchinson_2020} for a thermal image of current-driven metallic liners for a qualitative comparison.}
    \label{fig:mcbride_initial_perturbation}
\end{figure}

This zone-by-zone random perturbation seeds MRT structures via a mechanism similar to those that occur due to imperfections in the target structure in actual experiments. Targets exhibit surface roughness as well as “voids” and “inclusions” in the metal, all of which lead to non-uniform current density across the surface of the target \cite{Yu_2023}. The non-uniform current density drives an instability known as the electrothermal instability wherein portions of the target that are hotter exhibit an increase in electrical resistivity and a positive feedback of increased ohmic heating; this mechanism is believed to seed the later-time macroscopic magneto-Rayleigh-Taylor structures  \cite{Peterson_2013}. Figure~\ref{fig:mcbride_initial_perturbation} shows a representative realization of the random initial condition in FLASH. 

\begin{figure}
    \centering
    \includegraphics[width=0.75\linewidth]{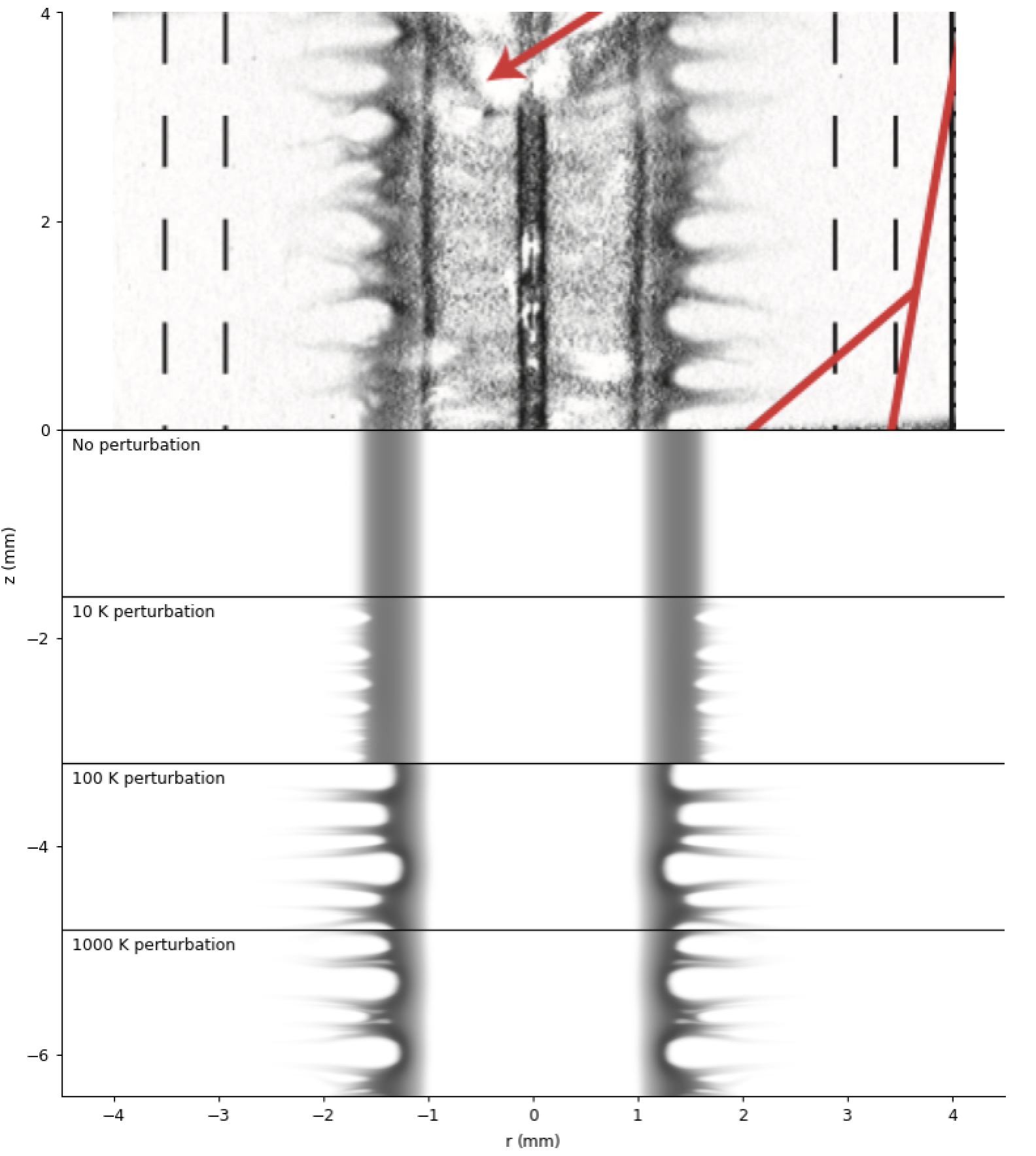}
    \caption{Radiograph of the multi-mode MRT structure at t=159.1 ns from \cite{McBride_2013} compared with FLASH simulations with random temperature perturbations ranging from 0 to 1000 K. Images from Ref.~\cite{McBride_2013} reproduced with the permission of AIP Publishing.}
    \label{fig:washington_radiograph_deltat}
\end{figure}

We performed a series of FLASH simulations with increasing temperature perturbation amplitude $\delta T$. The simulation domains had an axial length of 1.6 mm with periodic boundary conditions. All FLASH simulations shown for this benchmark use a finest AMR mesh resolution of 12.5 \micron and have radiation physics turned off for computational efficiency due to their negligible impact in this regime. A snapshot at time $t=159.1$ ns is shown in Fig.~\ref{fig:washington_radiograph_deltat}, comparing the Abel-inverted experimental radiograph with four simulations with $\delta T = 0, 10, 100$ and 1000 K. In the experiments, a thin layer of aluminum was coated on the liner inner surface for contrast which appears as a thick black line in the Abel inversion. We have not included this layer in our simulations because we can easily distinguish the liner inner surface in the simulation. Without the temperature perturbation, the FLASH results remain one-dimensional throughout the implosion, showing no noticeable MRT structure. This is encouraging, as it indicates that the code acceptably preserves symmetry in the axial direction in the absence of an initial perturbation. As we increase the perturbation, at 10 K we observe significantly less MRT structure than the experimental radiograph. Meanwhile, both the 100 K and 1000 K are in general qualitative agreement with the experiment. 

\begin{figure}
    \centering
    \includegraphics[width=0.9\linewidth]{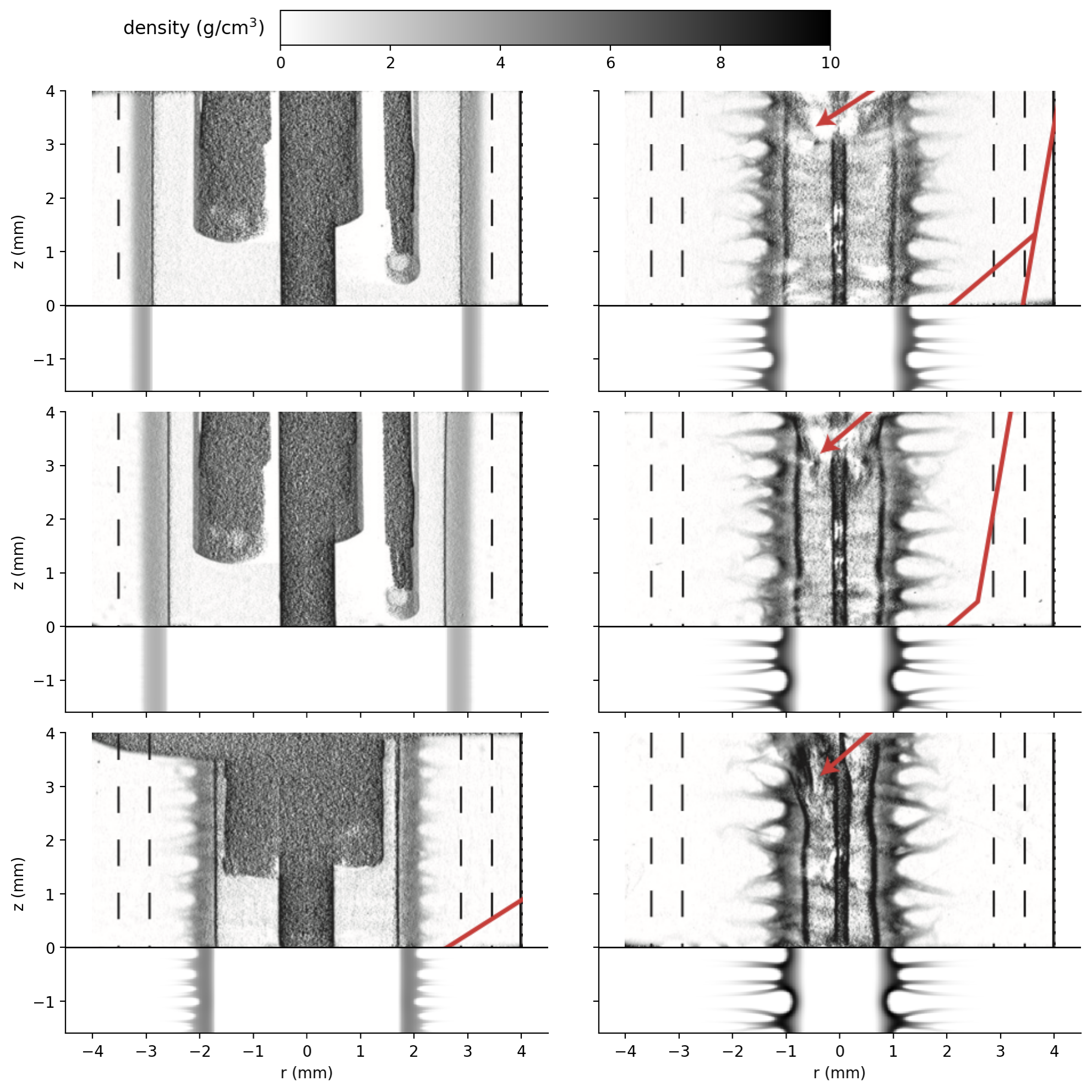}
    \caption{FLASH simulations compared to Abel-inverted radiographs taken at six instants in time: 100.4, 120.5, 149.8, 159.1, 162.1, and 164.9 ns. Radiograph data from Ref.~\cite{McBride_2013}. In each of the six panels, the experimental data is shown on top and the FLASH simulation is shown on bottom (z < 0). Images from Ref.~\cite{McBride_2013} reproduced with the permission of AIP Publishing.}
    \label{fig:washington_radiograph_time}
\end{figure}

Finally, we present the full time series comparisons against the experimental radiographs using the $\delta T = 100$ K FLASH simulation in Fig.~\ref{fig:washington_radiograph_time}. Simulation times are selected to match the liner convergence; the experimental data comes from several shots with enough variation in timing and magnitude of the current delivery so as to limit the utility of absolute timing comparisons. Other benchmarks in this document demonstrate FLASH’s ability to match experimental and simulated implosion trajectories including absolute timings. In this comparison, when matching the liner at similar convergence, we observe the prominent characteristics of the MRT features are in good agreement between the simulation and the experiment. This includes the amplitude of the largest mode, the appoximate wavelength of the largest mode, and general qualitative agreement in the macroscopic multimode behavior. In net, this benchmark validates that FLASH is capable of modeling the nonlinear evolution of MRT structures at the liner-vacuum interface. The fact that such a simple model for seeding the MRT structures does a decent job of capturing the late time MRT evolution is encouraging. The perturbation model could be refined further, but useful estimates of the effect of experimentally relevant MRT structures may apparently be made using this simple temperature perturbation model. 

\subsection{Benchmark 3: Single-Mode Richtmyer-Meshkov}
\label{ssec:single-mode_RM}

Here we validate FLASH’s ability to capture shocks and shock-induced hydrodynamic motion in magnetically driven implosions. We compare against experimental radiographs of shock-induced growth of sinusoidal perturbations on a beryllium rod at the axis of a cylindrical target implosion. FLASH accurately captures the experimentally measured growth rate of the shock-driven hydrodynamic mode in the linear regime.

The Richtmyer-Meshkov instability (RMI) is ubiquitous in implosion platforms such as those commonly found in ICF.  This instability occurs when a shock crosses an interface between two media of different densities.  The shock impulsively accelerates the interface and vorticity is generated where the shock interacts with perturbations at the interface.  In the post-shock flow, which is characterized by a constant velocity, this vorticity grows causing the perturbations to grow.  The perturbations will grow linearly in time until they reach an amplitude that is comparable to the wavelength of the perturbation.  After this point the growth continues but at a slower rate.  In a converging geometry, the situation is complicated because the shock and interface do not generally evolve with a constant velocity.  Additionally, as the perturbations approach the axis the gradient in pressure grows due to convergence effects.  This causes the bubbles and spikes to experience different environments, causing them to evolve at different rates, exacerbating the nonlinear behavior.  The “Decel” platform on Z was devised precisely to study this effect, as described in Ref.~\cite{Knapp_2020}.

\begin{figure}
    \centering
    \includegraphics[width=\linewidth]{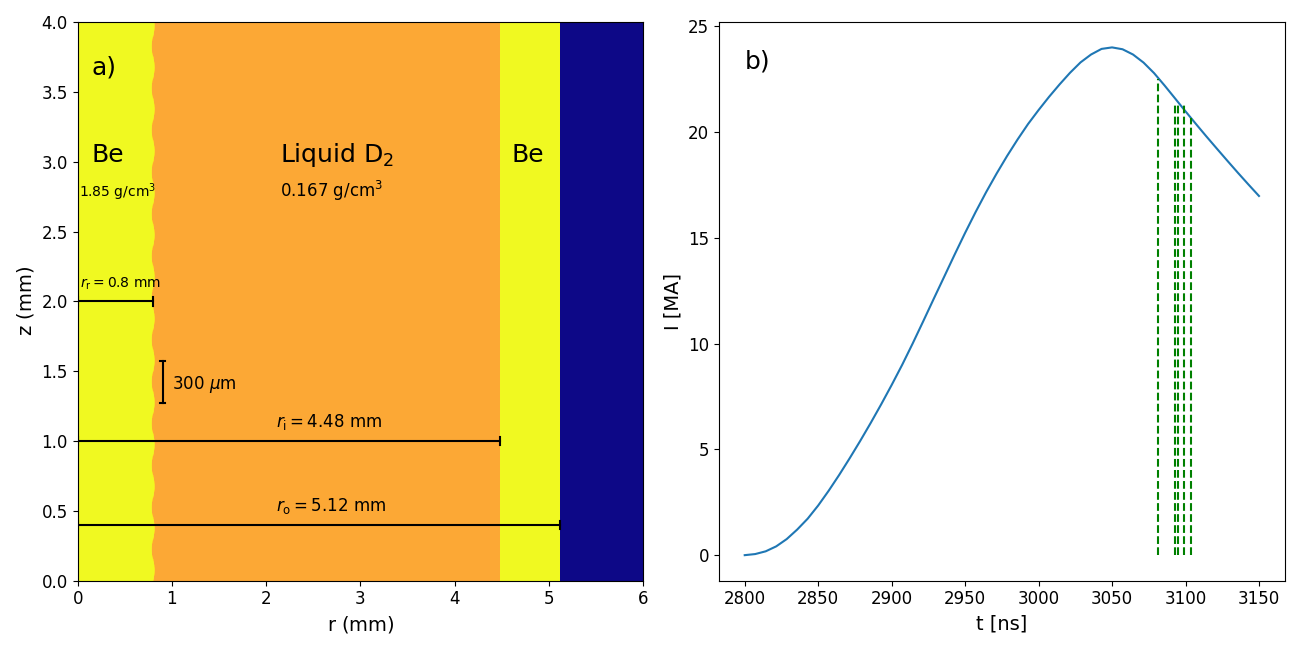}
    \caption{ a) Schematic of the target ~from Ref.~\cite{Knapp_2020} including the on-axis beryllium rod, liquid deuterium working fluid, and beryllium liner. b) Current trace for the RM platform experiments with times of radiograph shown in vertical lines.}
    \label{fig:decel_schematic}
\end{figure}

The Decel platform, shown schematically in Fig.~\ref{fig:decel_schematic} (a), uses an imploding beryllium liner to drive a converging hydrodynamic shock into a liquid deuterium working fluid.  The deuterium transmits the shock into a beryllium rod on the axis of symmetry.  This rod has a perturbation machined into it, typically a single mode sinusoidal perturbation with $k$-vector aligned along the $z$-axis.  After the shock impacts the beryllium/deuterium interface the rod compresses at a roughly constant velocity and the perturbation grows.  Figure~1c from Ref.~\cite{Knapp_2020}  shows the evolution of the experiment from a 1D ALGERA simulation.  The red lines show the location of the outer and inner surfaces of the liner as well as the evolution of the interface.  The superimposed gray scale image shows the gradient of the material pressure from the calculation.  This allows shocks, where there is an instantaneous jump in pressure, to show up prominently.  Figure 4 of Ref.~\cite{Knapp_2020} shows radiographically inferred densities obtained on these experiments at different instances in time; the growth of the perturbation is obvious in these images as well as the location of the shock in the rod.  

\begin{figure}
    \centering
    \includegraphics[width=0.5\linewidth]{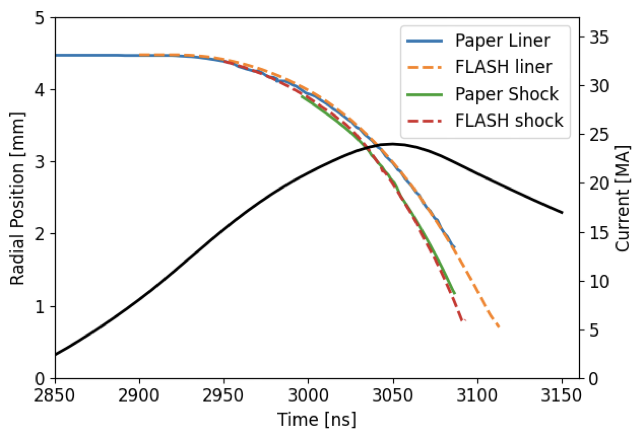}
    \caption{Comparison of the position of the inner liner radius and shock position in the FLASH simulation of the Decel platform to the positions inferred from the PDV diagnostic. The left y-axis shows radius while the drive current is shown in black and the right y-axis. The FLASH trajectory agrees very well with the experimental data.}
    \label{fig:decel_r_vs_t}
\end{figure}

In addition to providing valuable constraining information about the evolution of the RMI, this platform also provides numerous observable quantities that uniquely constrain a range of important quantities from the acceleration of a magnetically driven liner to the equation of state of beryllium and deuterium in regimes of interest for high-gain fusion.  The measurements of interest for the purposes of the validation of FLASH are the x-ray radiography and radial Photonic Doppler Velocimetry (PDV).  The PDV probes are placed on the axis of the cylinder, where the rod is.  The probes only take up a small portion of the axial extent of the rod, thus not interfering with the RMI growth measurements.  The PDV probe consists of fiber optics transmitting light radially outward, reflecting off of the inner surface of the liner, and then recollecting the light.  When the light reflects off of a moving surface it acquires a phase shift (similar to the Doppler effect) which can be detected via interferometry.  As the liner begins to implode PDV can measure only this velocity.  However, the motion of the liner soon generates a shock in the deuterium.  For a period of time early in the experiment the PDV probes can measure both signals simultaneously.  Eventually, the shock becomes opaque and we can no longer measure the liner velocity.  Finally, as the shock begins to approach the radius of the rod we lose the signal entirely.  The experimentally measured data is shown in Fig.~\ref{fig:decel_r_vs_t} aligned with the results from a FLASH calculation of the experiment.

We modeled the Decel platform in FLASH in two dimensions. The target dimensions are provided in Fig.~\ref{fig:decel_schematic} (a) and we simulated 1.8 mm axially (six wavelengths of the rod perturbation) and 8 mm radially with vacuum surrounding the beryllium liner. To seed MRT growth, we applied the random temperature perturbation initial condition of Eq.~\eqref{eq:tele_perturbation} with an amplitude of 40 K and decay length of 0.005 cm. We used five levels of AMR refinement with a resolution of dx=3.90625 \micron and dy = 3.515625 \micron at the finest levels (slightly non-square zones due to the aspect ratio of the problem).

The experimental data provides important information to constrain the energy coupling from the driver to the liner and the acceleration rate of the liner; while it does not directly inform the equation of state of the liner and the liquid fill materials, it is sensitive to them.  Figure~\ref{fig:decel_r_vs_t} shows a comparison of the liner and shock positions as measured in the experiment (solid blue and green, respectively) and simulated using FLASH (dashed orange and red). The FLASH inner radius was determined by calculating the average density in the axial direction and locating the position at which the average density exceeded 5\% of the maximum. For the shock position, we found the location of maximum $(d \log(p)/dr)$ The driver current is shown as the black line.  As we can see, there is excellent agreement between the measured and simulated trajectories.  This level of agreement, which is at least as good as that demonstrated in the paper between ALEGRA and the experiment (for comparison, see Figure~1 in Ref.~\cite{Knapp_2020}), indicates that we have a high level of confidence in our ability to model the imploding liner dynamics in magnetically driven targets.

To benchmark the behavior of the shocked on-axis rod and the Richtmyer-Meshkov evolution, we post-processed the FLASH simulation to obtain the amplitude growth of the perturbation on the rod for comparison to the experiment. In Fig.~\ref{fig:decel_mode_amplitude} below, we show the growth factor, defined as the instantaneous amplitude divided by the initial amplitude, vs. the so-called normalized interface position.  This quantity, which is the distance the interface has traveled at a given time multiplied by the wavenumber of the perturbation gives a sense of how far the interface has traveled relative to the scale of the perturbation. This in principle allows many different datasets to be compared on equal footing. The FLASH results are shown as the dashed blue line and the magenta circles are the experimental data. We see excellent agreement up until the last data point. In our simulations, the interface is reshocked sooner, halting the perturbation growth earlier than what is observed in the experiment.  We believe that this discrepancy is due to the EOS of the post-shock beryllium, leading to more compressibility of the rod in our simulations. In light of the good agreement up until this final data point, we conclude that FLASH is accurately capturing shocks and shock-driven hydrodynamic instabilities. 

\begin{figure}
    \centering
    \includegraphics[width=0.5\linewidth]{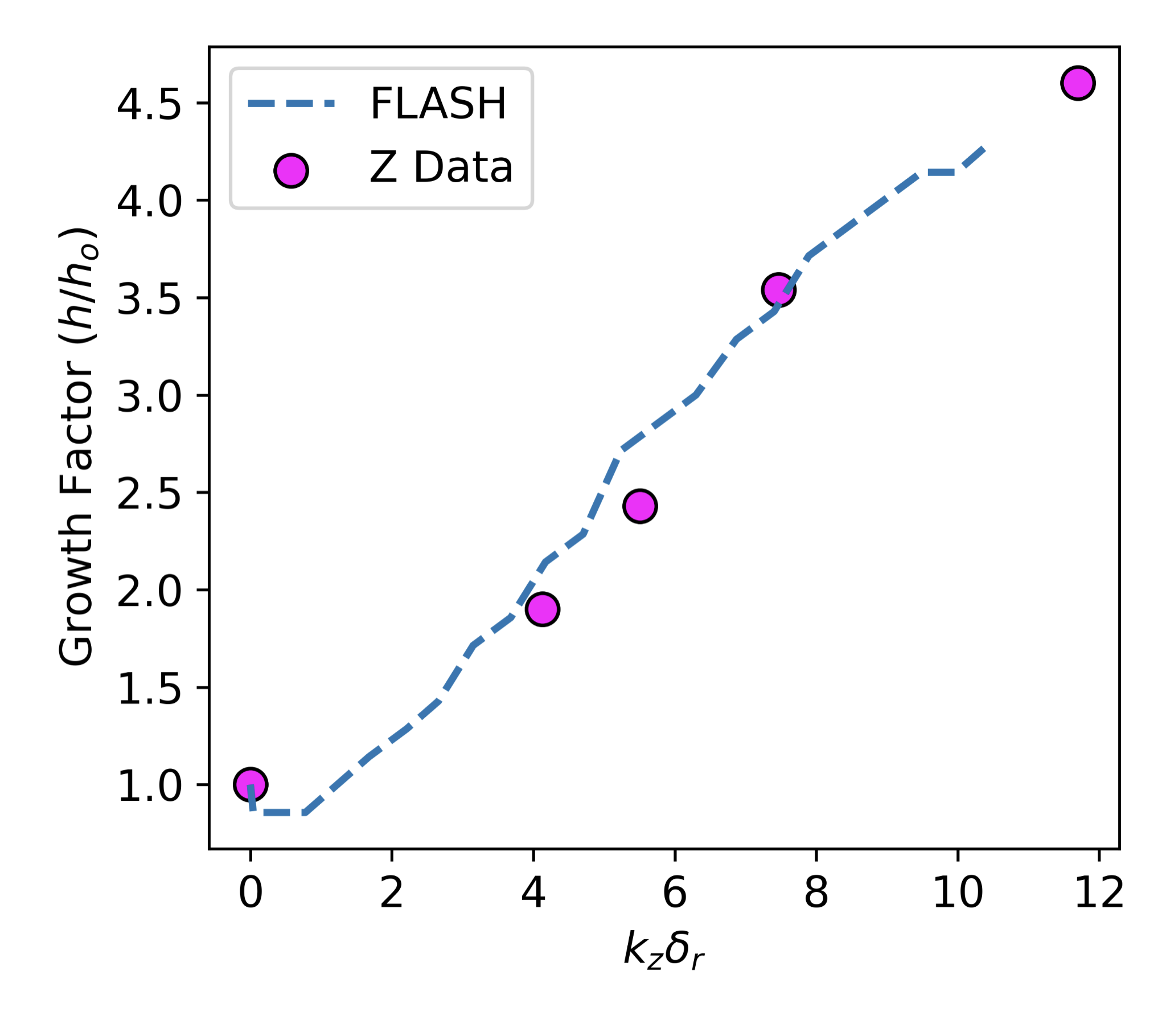}
    \caption{Growth of the Richtmyer-Meshkov instability in the experimental data from Ref.~\cite{Knapp_2020} and from 2D FLASH simulations. FLASH is recovering the correct linear growth rate, although it becomes re-shocked and enters the nonlinear evolution slightly earlier than observed in the experiment. }
    \label{fig:decel_mode_amplitude}
\end{figure}

\subsection{Benchmark 4: ICF Confinement Time}
\label{ssec:icf_confinement_time}

The inertial confinement time of the compressed fusion fuel is a key characteristic for fusion performance. To experimentally measure this important quantity, Knapp et al. developed a platform on Z involving a beryllium liner compressing liquid deuterium fuel~\cite{Knapp_2017} to diagnose the time evolution of the fuel near peak compression, revealing the hydrodynamic inertial timescale of the stagnated implosion. Early time evolution of the liner and shock breakout were measured using PDV probes. Late time convergence ratios and inertial confinement time were diagnosed using x-ray radiographs timed near peak stagnation for a collection of experiments nominally involving the same conditions. Simulations of the experiment using the ALEGRA code were performed in 1D and 2D and presented in Ref.~\cite{Knapp_2017}.

The target for the platform is a 3.84 mm outer-radius beryllium liner with an aspect ratio of 9.6 filled with liquid deuterium fuel of initial density 0.167 g/cm$^3$ . The current rises over $\approx$300 ns to a peak of about 12 MA. Over the course of the implosion the target compresses by a factor of $\approx$8 in radius leading to a $\approx$400 \micron radius compressed fuel with density $\approx$10 g/cm$^3$ at stagnation. 

\begin{figure}
    \centering
    \includegraphics[width=\linewidth]{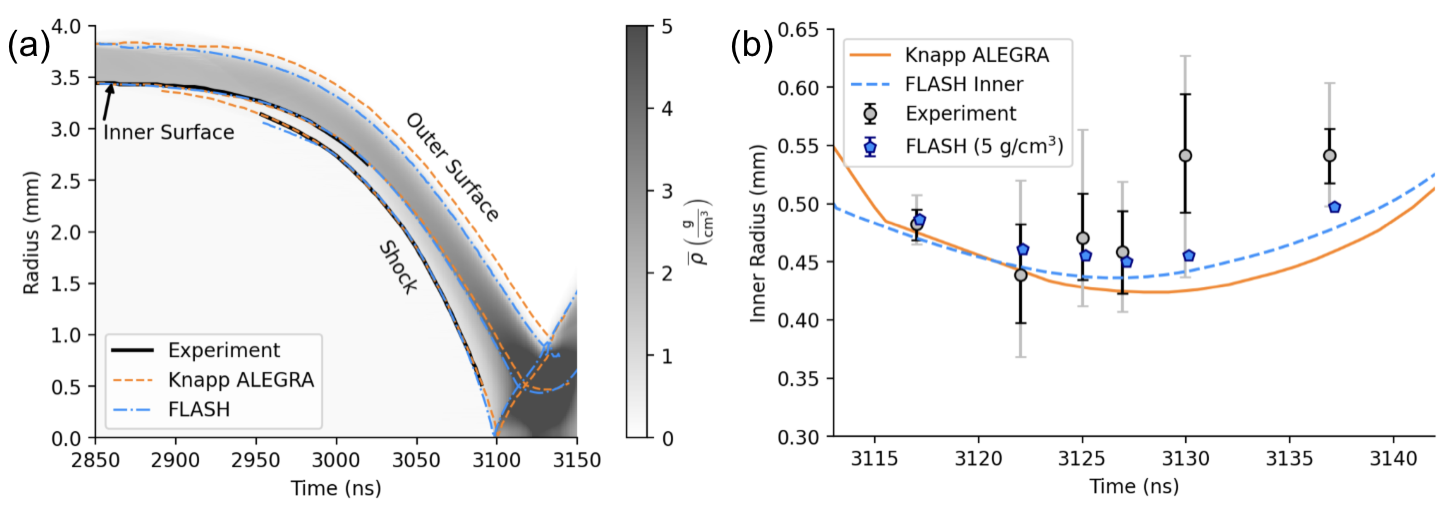}
    \caption{(a) 1D FLASH simulations of the ICF confinement platform compared to the PDV-inferred trajectory and the 1D ALEGRA simulation from Ref.~\cite{Knapp_2017}. (b) Stagnation-time dynamics of the inner liner trajectory for 1D FLASH, 1D ALEGRA and as inferred from the experimental radiographs according to Ref.~\cite{Knapp_2017}. }
    \label{fig:eddy_r_vs_t}
\end{figure}

In Fig.~\ref{fig:eddy_r_vs_t}, we show 1D FLASH simulations of this platform. These simulations were driven using a current source provided by the author of the publication and is similar to that shown in Fig. 2 of Ref.~\cite{Knapp_2017}. We used three levels of AMR with a finest resolution of 5.2 $\micron$. In panel (a), the FLASH trajectory shows excellent agreement with the inner-surface evolution and the shock timing in the fuel, agreeing with the experimental data and ALEGRA simulations to ns-scale precision. On the outer surface, no experimental data is available. The ALEGRA contour represents a Lagrangian tracer while the FLASH contour represents a constant density of 1.0 g/cm$^3$ so quantitative agreement is not expected but peak velocities appear similar. In panel (b), we show the dynamics near stagnation. The inner liner surface in FLASH reaches a minimum radius of 0.45 mm, in very good agreement with the experimental data and ALEGRA simulations; the timing of the stagnation is also well-aligned, confirming that FLASH is modeling the same ICF confinement time as experimentally observed. Finally, 5 g/cm$^3$ inner density radii are shown in FLASH as a representative comparison with the experimentally inferred inner radii from the radiographs. The 1D FLASH simulations overlap the error bars for the radiographic data.

\begin{figure}
    \centering
    \includegraphics[width=\linewidth]{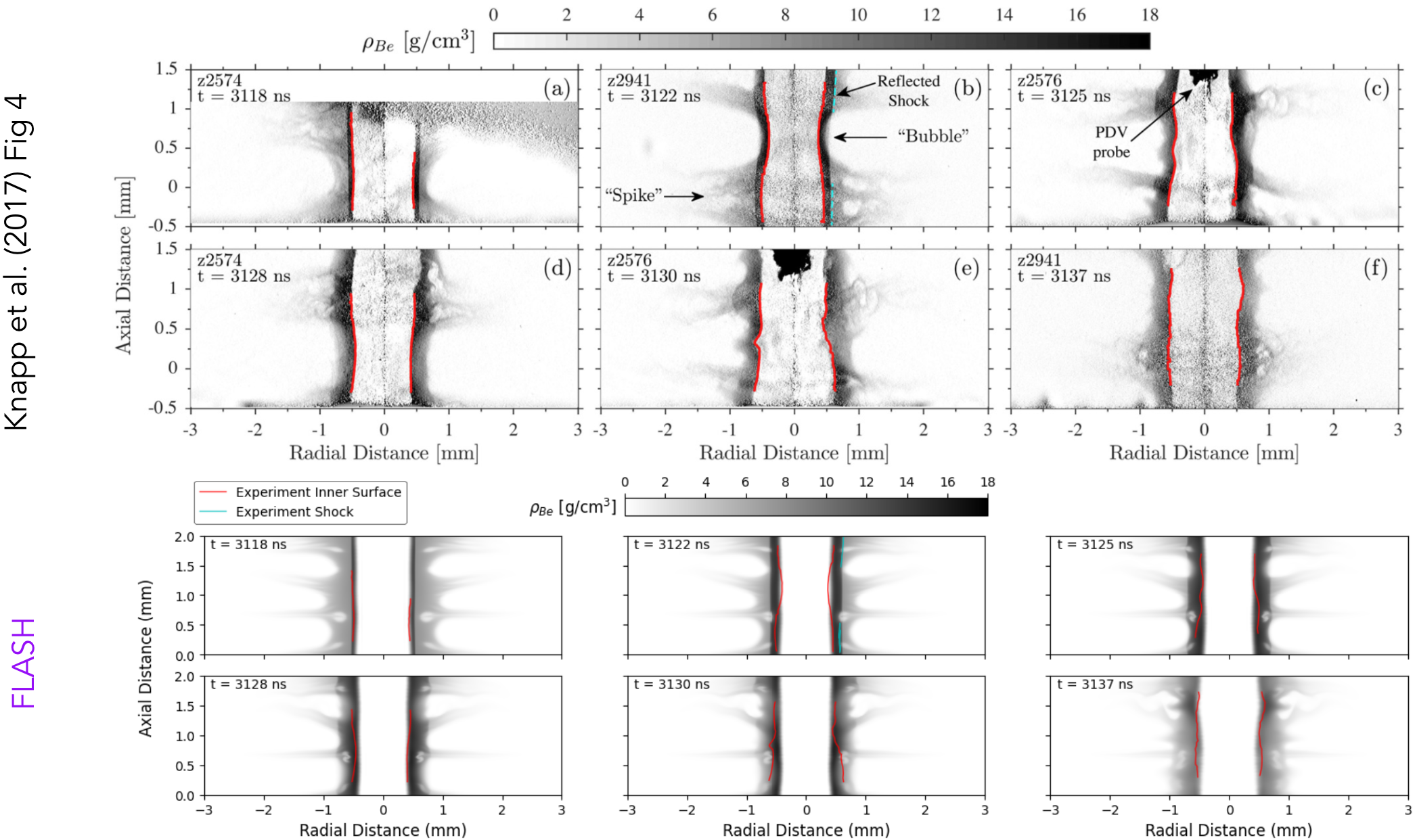}
    \caption{ Top - Mass densities inferred via Abel inversion of 6.151 keV x-ray radiographs of the beryllium liner near peak convergence (reprinted from Ref.~\cite{Knapp_2017} with the permission of AIP Publishing). Bottom - 2D FLASH simulations at the same times. }
    \label{fig:eddy_radiographs}
\end{figure}

Next, Fig.~\ref{fig:eddy_radiographs} shows 2D FLASH simulations compared to the densities inferred via Abel inversion from the experimental radiographs near stagnation. These simulations used three levels of AMR with a finest resolution of 15 um. We simulated 2 mm in axial extent with periodic boundary conditions. Many of the key characteristics are in good agreement between the simulation and the data including the inner liner radius, the general amplitude and dominant wavelengths of the MRT structures, and the timing of the outgoing shock which is visible as the dark boundary layer moving radially outward through the beryllium in time. The experiments show more prominent MRT structure along the inner surface; for instance at time t=3122 ns the FLASH simulation is essentially unperturbed while the experimental data shows a noticeable perturbation on the inner surface. These structures may be better captured with a different seeding mechanism for the MRT structures or they may relate to experimental imperfections not captured in the simulations. Overall, the agreement between FLASH and the data is similarly good as those obtained from the 2D ALEGRA simulations; see for instance Fig. 6 from Ref.~\cite{Knapp_2017} which also shows no MRT effect on the inner liner interface and arguably has less similar MRT structures at the vacuum/liner interface.

Overall, this benchmark enhances confidence in FLASH’s ability to capture stagnation-relevant performance quantities, at least in low to moderate convergence ratio implosions. The later benchmarks validate FLASH’s accuracy for implosions with higher convergence ratio and faster-rising current pulses.

\subsection{Benchmark 5: MagLIF Shot z2977}
\label{ssec:MagLIF_z2977}

Next, we benchmark FLASH against a fully integrated MagLIF experiment that has been described in multiple publications~\cite{Knapp_2019, Gomez_2020, Lewis_2021, Lewis_2023, Ampleford_2024}. This validates a large set of interacting physics, extending the previous validation benchmarks to include laser pre-heated fuel, the presence of externally imposed axial magnetic fields which must evolve hydrodynamically and correctly influence the thermal evolution of the hot fuel, and the measurable presence of DD fusion reactions. The experimental data consists primarily of measurements of the fusion-generated neutrons from which one can infer the energy yield, the temperature of the plasma ions undergoing fusion, and the pressure of the fusion fuel averaged throughout the rapid burn process. Accurately simulating these quantities is a stringent test on the integration of the many underlying physical processes. We find that FLASH agrees well with the experimental data, exhibiting comparable accuracy to the LASNEX multiphysics code which is regularly used by Sandia to design MagLIF targets.

Z shot z2977 is a beryllium liner target with beryllium cushions, aspect ratio 6, inner radius 2.325 mm, filled with 0.68 mg/cm$^3$ of deuterium gas, magnetized with 10 T of axial magnetic field, and preheated with 0.775 kJ of laser pre-heat (note that the analysis of inferred preheat deposition evolved throughout the publication history; we defer to the more recent publications as the best assessments). It produced a nuclear yield of $3\times10^{12}$ neutrons with a burn-weighted ion temperature of $\approx$2.6 keV. The extensive analysis performed on this shot across the publications together with published LASNEX simulation results made it a useful focus for our Z-scale MagLIF FLASH benchmark.

\begin{figure}
    \centering
    \includegraphics[width=0.5\linewidth]{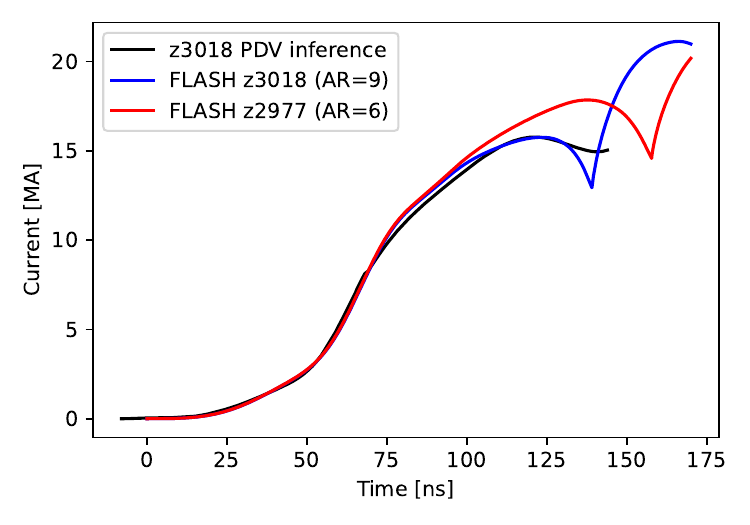}
    \caption{Comparisons of simulated and measured load currents for AR=9 Z shot z3018. The load current was inferred from photonic doppler velocimetry measurements on the transmission line anode (``return can”). The circuit settings for the 1D FLASH simulation of z3018 were applied to the thicker but otherwise identical liner z2977, which implodes more slowly and reaches a higher peak current. }
    \label{fig:z977_current_benchmark}
\end{figure}

To model this experiment, we used an external, lumped-element circuit model to provide magnetic energy to the MHD simulation domain while receiving dynamic feedback from the time-varying impedance of the load. Details of the circuit model implementation are provided in Section~\ref{sec:background} and validation of its behavior in modeling MagLIF systems is provided in Benchmark 6. For z2977, the voltage-driven circuit model was calibrated to a recently published analysis of load current for a MagLIF experiment \cite{Hutchinson_2023}. This experiment z3018 tested a thinner AR=9 liner than the AR=6 liner used in z2977. The circuit used is similar to that shown later in the circuit diagram in Figure 1 of Ref.~\cite{Ruiz_2023_Iscaling} but with changes to several parameters to reflect typical values used to match Z experiments. In particular, the loss resistor timing and rate-dependence was varied to achieve the match to the z3018 inference. This circuit model was then applied to the AR=6 z2977 liner simulations. The quoted current inference for z3018 was $I_\mathrm{max} = 16 \pm 0.2$ MA, but the author admits this uncertainty is an underestimate; $\pm$ 1 MA is a more typical estimate for Z experiments. This uncertainty should be considered when comparing simulated stagnation conditions and thermonuclear outputs to measurements. The resulting current drive from this calibration is shown in Fig.~\ref{fig:z977_current_benchmark}.

Additional physics included in FLASH simulations of z2977 include a pre-heat energy source, Nernst advection of magnetic flux, and hydrogenic thermonuclear burn. The pre-heat energy source deposits a prescribed amount of energy in a volume of fuel of fixed radius and axial extent over some user-defined time window. Although FLASH offers a laser ray trace package for high fidelity deposition of laser energy, we chose to bypass the additional complexity and instead directly deposit the desired amount of energy, a practice which is often used when planning MagLIF experiments at Z. Nernst extended MHD is an important effect for advecting magnetic flux away from the hotspot due to the thermal gradients that emerge during the laser pre-heat. Various MagLIF publications have shown sensitivity with respect to this effect and we have included it in the FLASH simulations. Finally, although the publicly distributed FLASH code does not contain TN burn for hydrogenic species (only higher Z fuel cycles relevant for astrophysics), we have extended our internal version of FLASH to perform TN burn, calculate burn-weighted metrics, and in the case of DT simulations optionally deposit alpha energy via an alpha energy diffusion model. 

\begin{figure}
    \centering
    \includegraphics[width=\linewidth]{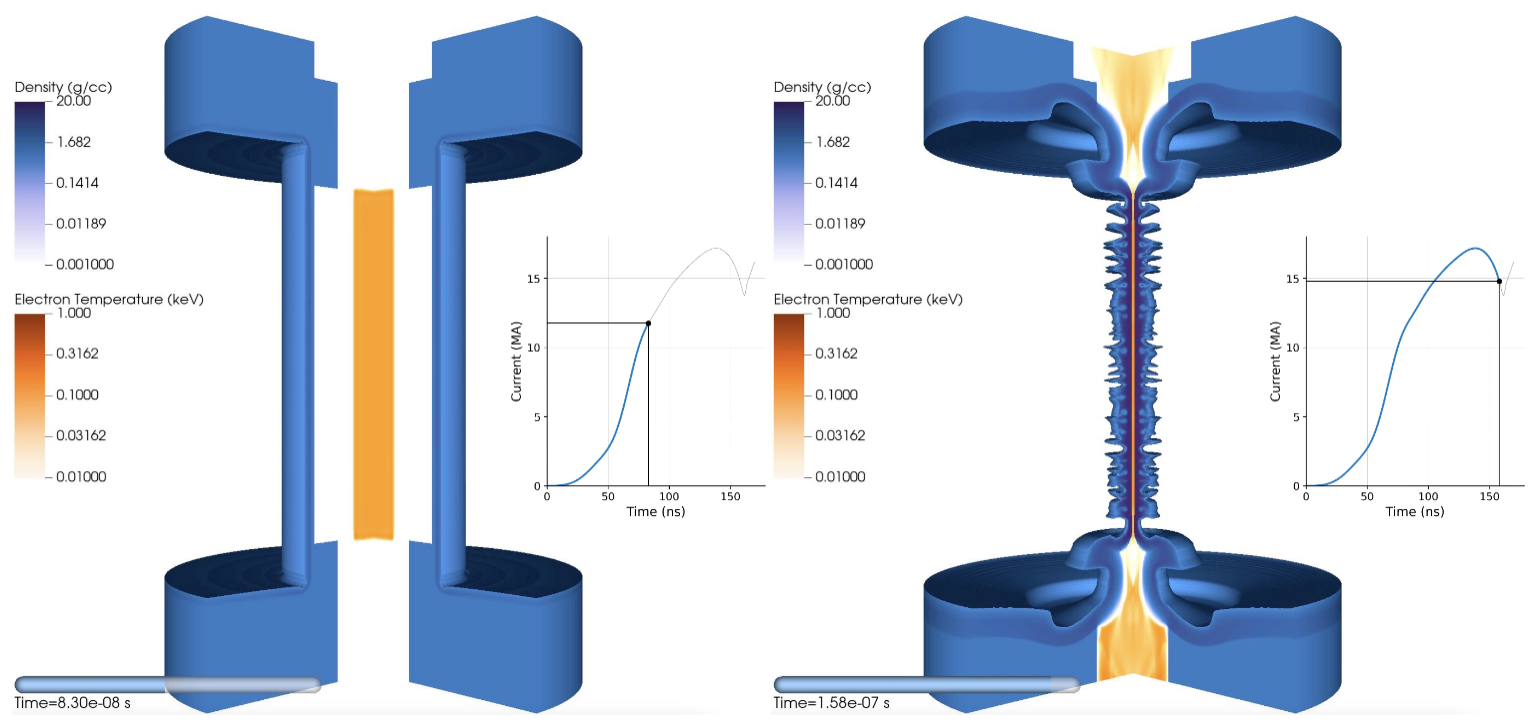}
    \caption{FLASH simulation snapshots of z2977 showing the state shortly after the pre-heat turn-on (left) and near peak compression (right). We revolve the 2D simulation in 3D for visualization and show the mass density in the beryllium liner and electrode regions and the electron temperature in the fuel region. }
    \label{fig:z2977_simulation_revolved}
\end{figure}

We ran 2D FLASH simulations of the full-scale target including cushion geometry and a portion of the radial electrodes, with the simulation domain extending 1.8 cm axially and 0.6 cm radially. The gap between the electrodes was 1.2 cm. Snapshots of the simulation are shown in Fig.~\ref{fig:z2977_simulation_revolved} and nuclear performance reported in Table~\ref{tab:z2977_performance}.  The simulations were run with seven levels of allowed mesh refinement corresponding to a highest resolution of 1.5 \micron. The nuclear performance indicates strong agreement between the experiment and 2D LASNEX simulations; the FLASH yield is $2.2\times 10^{12}$ neutrons, which is slightly below the experimental yield of $3\times10^{12}$ neutrons. Many factors could affect this agreement including details of the preheat treatment, mesh and temporal convergence and material properties. The LASNEX simulations report a yield of $6\times10^{12}$ neutrons - one takeaway being that the level of accuracy obtained by FLASH is adequate for performing target design with a similar predictive capacity as the standard design codes used at the national labs.

\begin{table}[h]
    \centering
    \begin{tabular}{lccc}
    \toprule
        & z2977 & LASNEX 2D &  FLASH 2D \\
    \midrule
        N Neuts & $3\times 10^{12}$ & $6\times 10^{12}$ & $2.2\times 10^{12}$ \\
        Tion [keV] & $2.63 \pm 0.53$ & 2.0 & 2.8 \\
        \bottomrule
    \end{tabular}
    \caption{Nuclear performance of Z shot z2977 compared with LASNEX simulations from Ref.~\cite{Gomez_2020} and our FLASH benchmark.}
    \label{tab:z2977_performance}
\end{table}

Data from z2977 were used to infer the magnetization level of the fuel BR, defined as
\begin{equation}
    BR = \frac{\Phi_z}{\pi R} = \frac{2}{R} \int_0^R r B_z dr
\end{equation}

\begin{figure}
    \centering
    \includegraphics[width=0.5\linewidth]{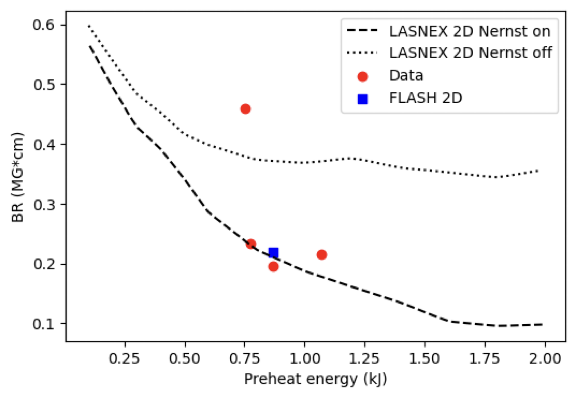}
    \caption{$BR$ at peak neutron production time from LASNEX 2D simulations \cite{Lewis_2021_ICOPS}, FLASH, and data-based inferences. 2D FLASH, which includes Nernst effects, agrees with 2D LASNEX and with data. }
    \label{fig:z2977_rb}
\end{figure}

Here $B_z$ is the axial magnetic field strength and $R$ is the radius of the fuel/liner interface. Publications from Sandia compare LASNEX calculations of $BR$ versus the laser preheat energy to the data-based inference \cite{Lewis_2021_ICOPS, Lewis_2021}. Figure~\ref{fig:z2977_rb} reproduces the LASNEX calculations and several data points from this paper. We add $BR$ at the time of peak neutron generation from the 2D FLASH calculation of z2977 to the plot. The FLASH 2D simulation with Nernst advection turned on agrees with the LASNEX calculation and the data for z2977, showing that FLASH can correctly calculate the diffusion and advection of the axial magnetic field in the fuel plasma. 

\subsection{Benchmark 6: MagLIF Current Scaling Study}
\label{MagLIF_current_scaling}

For our final benchmark, we reproduce the MagLIF current scaling study of Ref.~\cite{Ruiz_2023_Iscaling}. This study scales a prototypical MagLIF target design from Z scale up to 60 MA while holding several dimensionless parameters constant to retain comparable hydrodynamic behavior across the study. The exercise is a strenuous test of physics capabilities in a current-driven ICF target design code including: dynamic coupling to an external circuit, in-plane components of the magnetic field $(B_z, B_r)$, a pre-heat energy source, Nernst extended MHD effects, and a sufficiently high fidelity model for alpha particle feedback on the hydrodynamic evolution. At the larger values of current in the study the fuel undergoes significant self-heating, exceeding a generalized Lawson ignition parameter $\chi = 1$ above $\approx$45 MA. The scaling relations were obtained in Ref.~\cite{Ruiz_2023_Iscaling} analytically with supplementary data using fully integrated 2D HYDRA simulations. The latter allows us to compare FLASH to HYDRA for high-performing MagLIF targets in current regimes relevant to Pacific Fusion's Demonstration System. 

\begin{figure}
    \centering
    \includegraphics[width=0.5\linewidth]{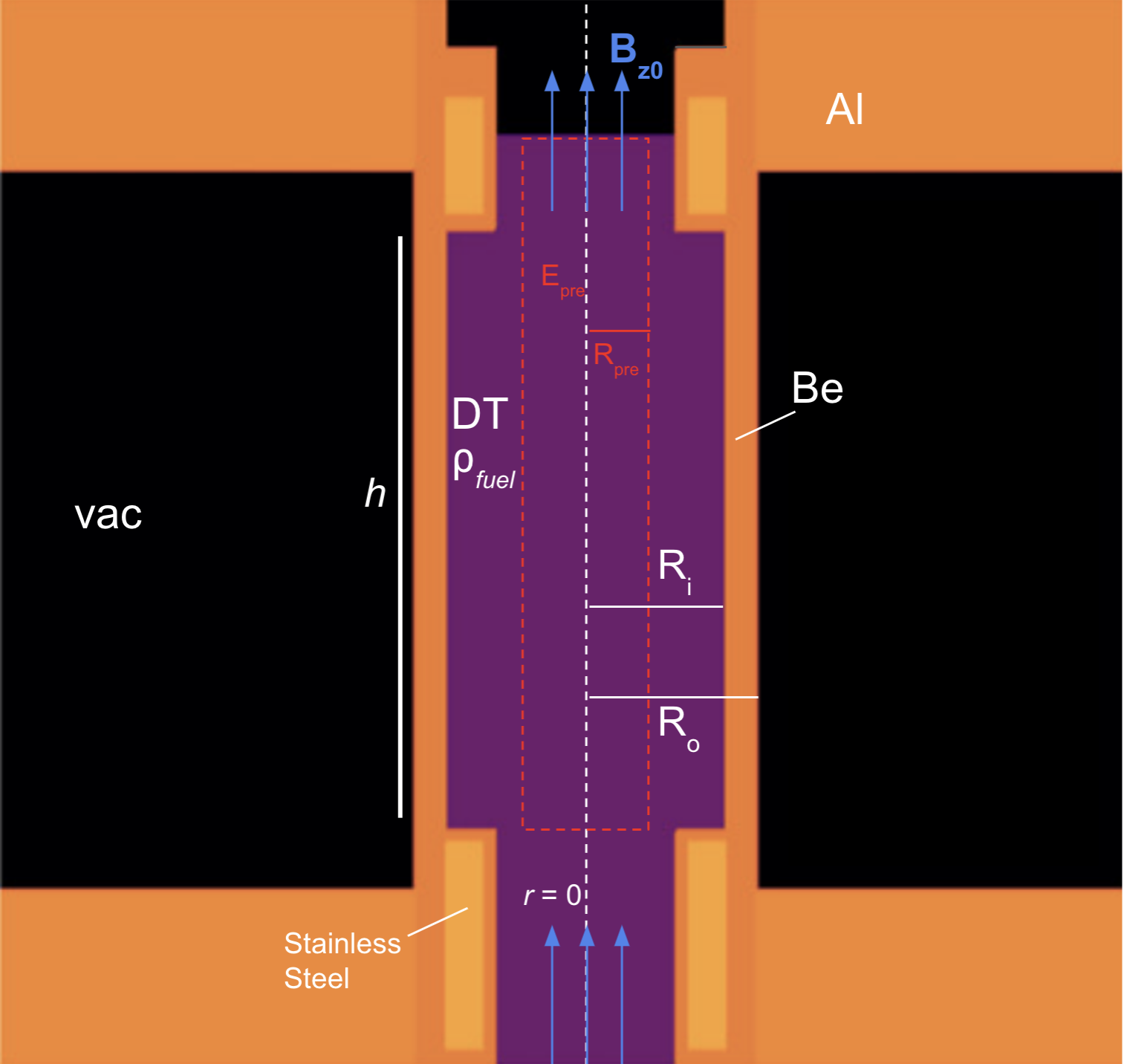}
    \caption{Schematic of the 2D MagLIF target used in the current scaling study of Ref.~\cite{Ruiz_2023_Iscaling}. }
    \label{fig:scaling_schematic}
\end{figure}

The MagLIF current scaling study in Ref.~\cite{Ruiz_2023_Iscaling} parametrizes a MagLIF target design via a set of independent parameters, then provides a prescription for scaling those parameters in such a way as to maintain constant values of key dimensionless parameters characterizing the implosion at all levels of current in the scaling study. A schematic of the target is shown in Fig.~\ref{fig:scaling_schematic}. The simulation includes aluminum electrodes, a beryllium liner, rectangular cushions at the four corners of the MagLIF target with stainless steel cores (these impede the wall mode bubble that forms at the corner of the electrode and the target), pre-magnetization with uniform axial magnetic field throughout the problem, DT fuel, and laser preheat. The key independent parameters influencing the fusion performance of the MagLIF target in this study are the liner height $h$, inner liner radius $R_\text{i}$, outer liner radius $R_\text{o}$, fuel mass density $\rho_\text{fuel}$, initial axial magnetic field $B_z$, preheat energy $E_\text{pre}$ and preheat radius $R_\text{pre}$. 

To obtain scaling relations for the independent parameters as a function of the peak current $I_\text{max}$, the scaling study considers a reference configuration of a target at $I_0 = 20$ MA, then scales the target parameters relative to those in the reference configuration in order to maintain several dimensionless parameters at all levels of current. For instance, the ratio of characteristic magnetic energy delivered to the liner to the characteristic kinetic energy of the liner is held constant, as is the susceptibility of the target to hydrodynamic instabilities and the rise time of the current pulse. For details of the scaling prescription, see Ref.~\cite{Ruiz_2023_Iscaling}. Ultimately, power-law relationships are provided for the independent target parameters as relative to their baseline values:

\begin{align}
    R_\text{i}' & = R_\text{i}\left( \frac{I_\text{max}}{I_0} \right)^{0.206} \\
    R_\text{o}' & = R_\text{o}\left( \frac{I_\text{max}}{I_0} \right)^{0.381} \\
    B_{z0}' & = B_\text{z0}\left( \frac{I_\text{max}}{I_0} \right)^{0.647} \\
    \rho_{\text{fuel}}' & = \rho_{\text{fuel}}\left( \frac{I_\text{max}}{I_0} \right)^{0.529} \\
    E_{\text{pre}}' & = E_{\text{pre}}\left( \frac{I_\text{max}}{I_0} \right)^{2.529} \\
    h' & = h\left( \frac{I_\text{max}}{I_0} \right)^{0.529}
\end{align}

For the circuit drive, the MagLIF target is driven by a lumped-element circuit model shown in Figure 1 in Ref.~\cite{Ruiz_2023_Iscaling}. The circuit includes a voltage source and a network of resistors, inductors, and a capacitor. The voltage source is tabulated in time to be representative of the Z driver with the baseline voltage trace shown in Figure 2 of Ref.~\cite{Ruiz_2023_Iscaling}. The voltage is scaled as the product of the height and the current scaling, so $\left(I_\text{max} / I_0\right)^{1.529}$, while the resistors and inductors scale linearly with the height; the capacitance scales inversely proportional to the height. Baseline values of the parameter study are provided in Table 3. Dimensions of the cushion geometry and electrode spacing are scaled linearly with $h$ and the laser preheat is deposited over 10 ns starting at 80 ns.

The MagLIF scaling study has been simulated in FLASH in one and two dimensions according to these prescriptions. An important consideration was the role of numerical mixing at the liner-fuel interface in influencing the nuclear burn process. We found that at coarse resolution the beryllium liner material mixed excessively into the fuel and suppressed the nuclear burn. A convergence study of 1D simulations found that 1.5 \micron mesh resolution was well converged in the nuclear burn metrics, so the 1D and 2D simulations were completed at this resolution. We used six levels of adaptive mesh refinement to achieve this.  

To benchmark the FLASH results, we compare several quantities from the referenced publication. In 1D, we compare the radial implosion trajectory of the liner to those presented for HYDRA and we compare the temporal current traces obtained by the two codes, which tests the feedback between the time-evolving target impedance and the external circuit drive. In 2D, we compare pseudocolor plots of the density at the initial condition and near peak compression. In both 1D and 2D, we compare the nuclear burn metrics of yield, burn-averaged ion temperature, burn-averaged pressure, and burn-averaged density, including cases with both alpha particle energy deposition turned on and off. When the alpha-on and alpha-off cases diverge, this indicates the onset of ignition. 

\begin{figure}
    \centering
    \includegraphics[width=0.8\linewidth]{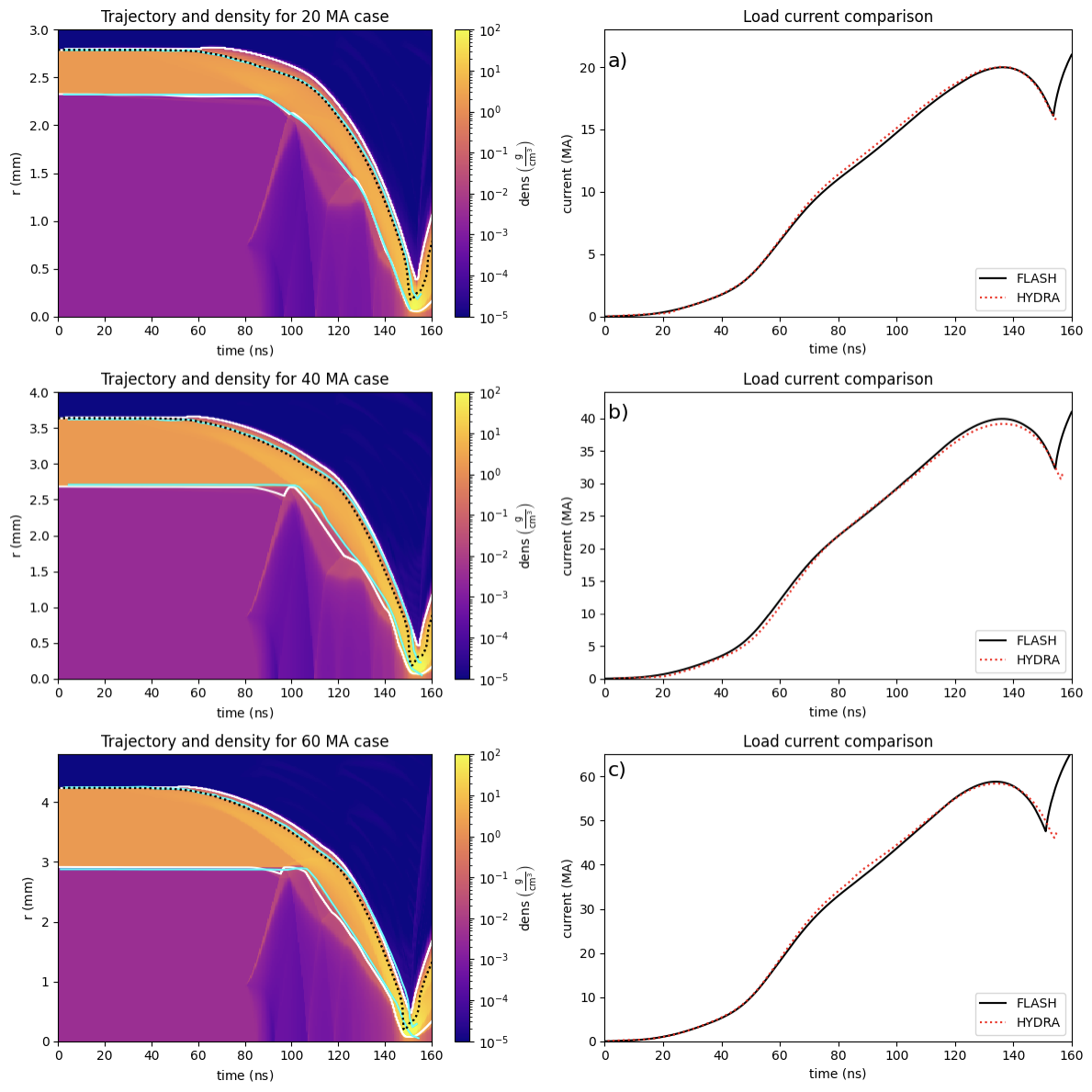}
    \caption{Comparisons of simulated trajectories (left) and load currents (right) at $I_\text{max} = $ a) 20 MA, b) 40 MA, and c) 60 MA.  White iso-contours of beryllium mass-fraction $m_\text{be} = 0.99$ show the edges of the beryllium liner. The dotted black line is the $1/e$ point of the outer liner density profile. HYDRA inner- and outer-liner trajectories are shown as cyan contours.}
    \label{fig:scaling_rho_t}
\end{figure}

First, Fig.~\ref{fig:scaling_rho_t} presents 1D FLASH simulation results at 20 MA, 40 MA, and 60 MA. In the left column, we show the mass density as a function of time and radius in pseudocolor. In these implosion trajectories we overlay several contours for tracking the positions of the inner and outer radii. For FLASH, the contour of 99\% beryllium mass fraction is shown in white and the position of the $1/e$ of the initial mass density is shown in dotted black. Meanwhile, contours of interface position for HYDRA calculations from Ref.~\cite{Ruiz_2023_Iscaling} are shown in cyan. The inner HYDRA contours represents a Lagrangian tracer and follows a similar evolution as the FLASH mass fraction contour at the inner radius. The outer HYDRA contour is a $1/e$ density contour which can directly be compared with the dotted black contour for FLASH, revealing good agreement. Important details to note are the implosion timing, including in-flight velocities and peak compression, and the amount of compression of the liner throughout the implosion. In the right column of Fig.~\ref{fig:scaling_rho_t} the load currents in the FLASH simulations agree quite well with those of the HYDRA simulations. Matching the simulated load current requires matching the outer liner trajectory. The load inductance increases as the liner implodes, causing the current to reach its maximum value followed by the “inductive dip.” 

\begin{figure}
    \centering
    \includegraphics[width=0.5\linewidth]{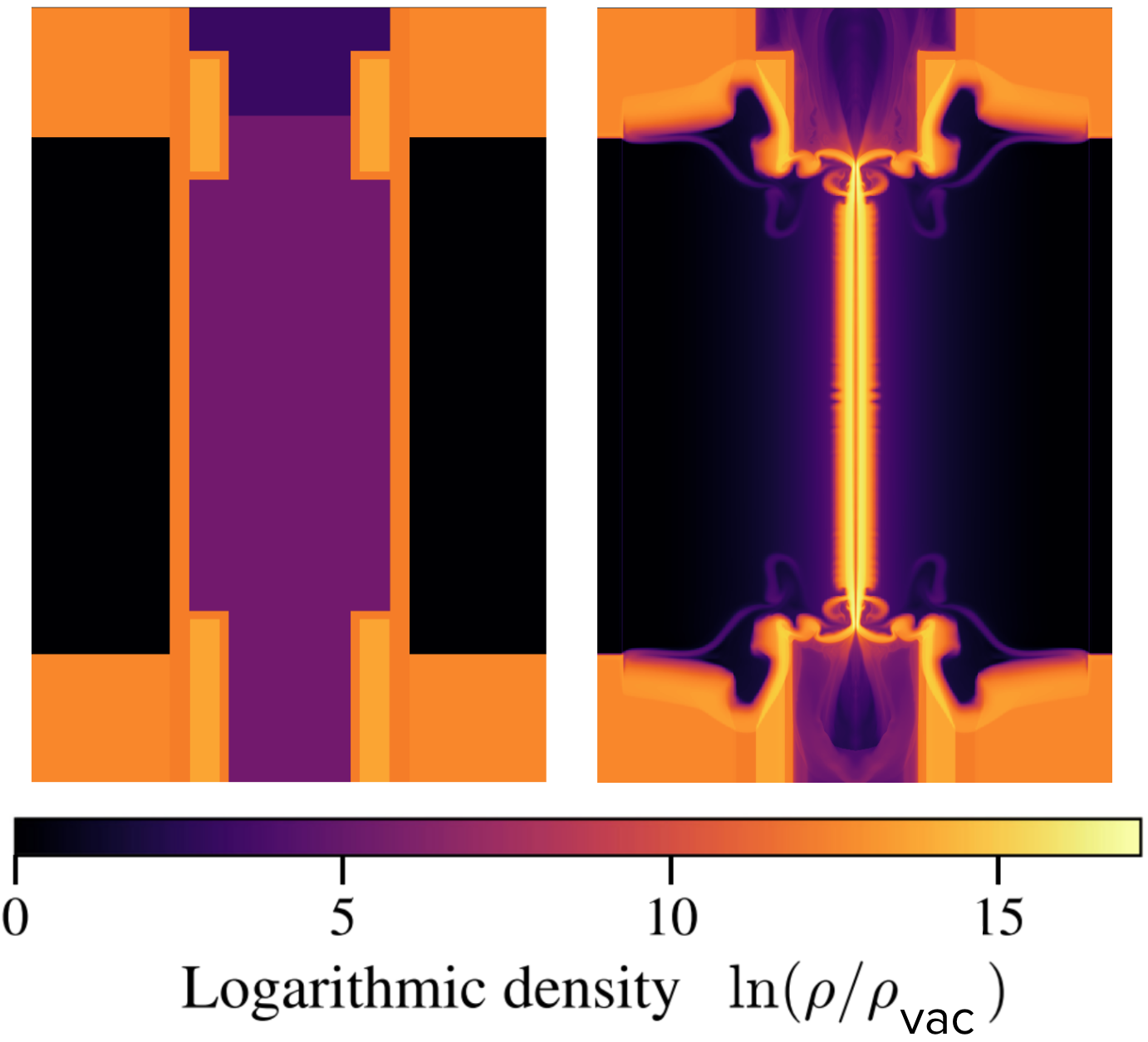}
    \caption{Mass density in FLASH 2D simulations of the 20 MA baseline point design in the current scaling study. The initial condition is shown on the left and a time near peak compression is shown on the right. The color bar shows the logarithmic density normalized by the vacuum density of $10^{-5}$ g / cm$^3$. For a similar figure with HYDRA, see Figure~5 in Ref.~\cite{Ruiz_2023_Iscaling}.}
    \label{fig:scaling_pseudocolor}
\end{figure}

Next, we present the 2D simulation results for this benchmark. Figure~\ref{fig:scaling_pseudocolor} shows pseudocolor plots of density at the initial time $t=$0 and at a time near peak compression. The time of the HYDRA image is not provided in the reference and the FLASH checkpoint files were only output at certain intervals so the timing cannot be taken to be exact in this comparison. Nevertheless, this qualitative comparison indicates that FLASH is obtaining a similar convergence ratio, exhibits similar dynamics at the corner of the electrodes and the liner, and achieves similar amounts of liner compression. Key differences in the comparison include the wavelengths of the MRT structures near peak compression; HYDRA appears to have longer wavelength and perhaps larger amplitude MRT structure while FLASH is generally more uniform with some small wavelength structures visible near the midplane. 

Finally, we compare the nuclear performance of FLASH to HYDRA across the range of currents represented in the study. Figure~\ref{fig:scaling_fusion_metrics} below shows fusion yield, burn averaged ion temperatures, burn averaged pressures, and burn averaged fuel densities. The burn averaged quantities are calculated according to e.g.
\begin{equation}
    T_\text{b,avg} = \frac{\int n_\text{DT}(r,z,t) T_\text{DT}(r,z,t) r \,dr\,dz\,dt}{\int n_\text{DT}(r,z,t) r\,dr\,dz\,dt}
\end{equation}
where $n_\text{DT}$  is the number of DT reactions taking place at a given position and time. The results ignoring the alpha particle energy deposition are shown in blue while those including alpha energy deposition are shown in red.

\begin{figure}
    \centering
    \includegraphics[width=0.9\linewidth]{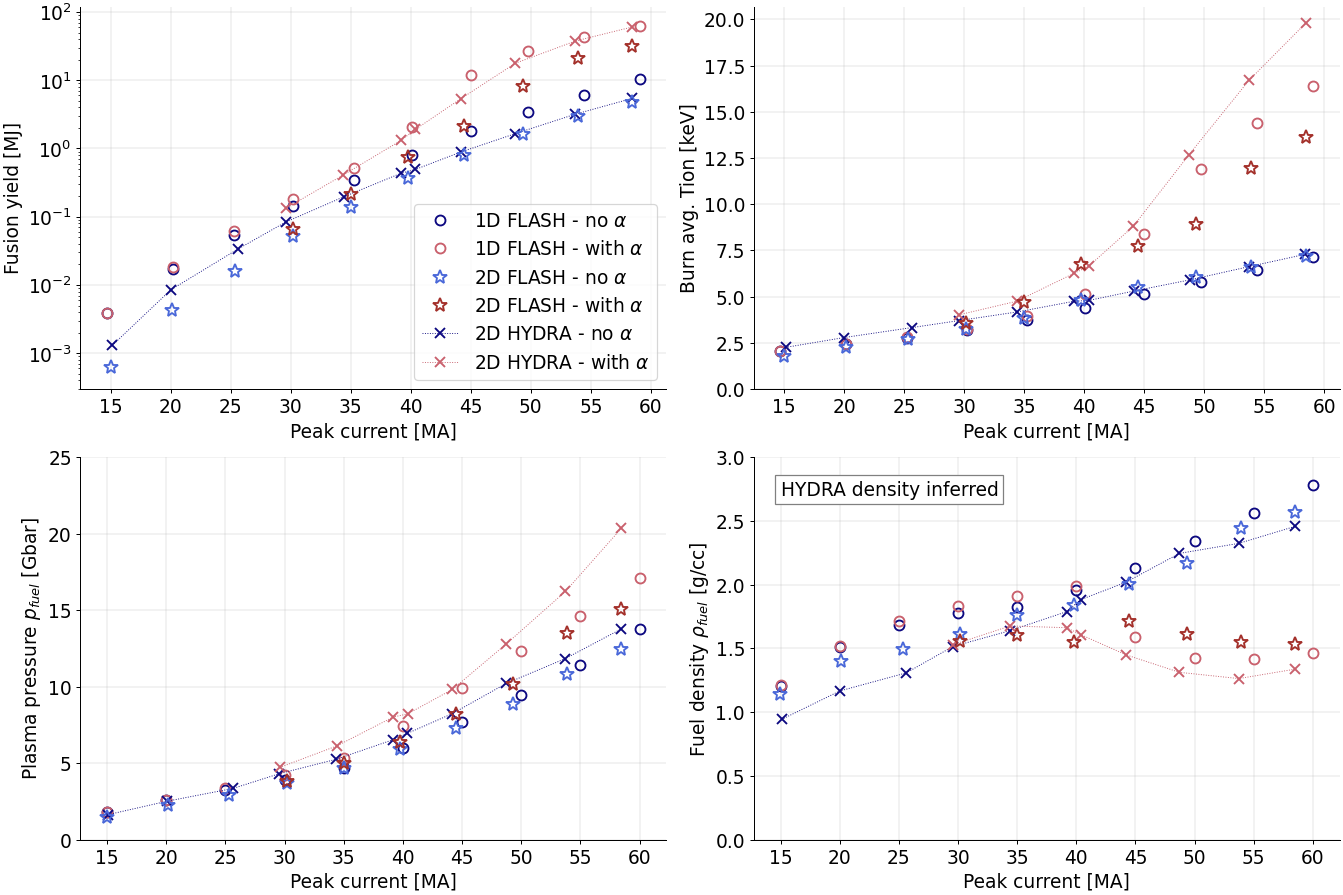}
    \caption{Fusion performance metrics for the current scaling study, including 2D HYDRA results from Ref.~\cite{Ruiz_2023_Iscaling} and 1D and 2D FLASH results. The temperature, pressure, and fuel density all represent burn-averaged quantities. The burn-averaged density is not reported in the reference, so we inferred the result using an ideal gas assumption and the burn-averaged temperatures and pressures.}
    \label{fig:scaling_fusion_metrics}
\end{figure}

Starting with the 1D FLASH yields, we find the results to be in reasonable agreement with HYDRA, with alpha-off yields being over-estimated by about a factor of 2 throughout the scaling and the alpha-on yields agreeing to within tens of percent for most points in the study. Over-estimating the yield in 1D is to be expected given the absence of end losses and any fuel compression asymmetries in the axial direction, e.g., due to magneto-Rayleigh-Taylor instabilities. For results including deposition of the alpha energy, an important consideration is the different models employed in transporting the alpha particles. The 1D FLASH results capture both qualitative trends, reflective of capturing the correct essential physics, and show surprisingly good quantitative agreement in the nuclear yields, indeed to an unexpected degree. One would expect a 1D HYDRA simulation to over-perform relative to the 2D simulation, so the close agreement between 1D FLASH and 2D HYDRA in the presence of alpha deposition is not necessarily a representation of exact correctness in the FLASH result. 

The 2D FLASH yields also agree with HYDRA to a satisfactory degree. For alpha deposition off, at low currents the yields are lower in FLASH but agree to within about a factor of 2. At high currents the yields agree very well. The overall trend in FLASH is quite similar to that exhibited in HYDRA, which indicates good agreement with the expected scaling behavior in the study. The alpha-deposition-on yields are also within a factor of 2 or 3 with the maximum discrepancy at intermediate levels of current, near $\approx$40 MA. This indicates a slight shifting of the fusion gain curve, which is likely due to differences in the alpha transport models. The alpha-on yields are lower in FLASH than in HYDRA, indicating that FLASH is making a more conservative prediction of fusion performance with the present collection of modeling choices. This is an important consideration for making confident predictions about ignition-scale targets; if FLASH were over-performing relative to HYDRA to a significant degree, one would be concerned that FLASH's ignition simulations may be overly optimistic, but this is not the case. Moreover, the presence of the ignition cliff as a function of current is significant, the specific location of which may be especially sensitive to modeling specifics.

Assessing the burn metrics of the hydrodynamic state $(T, P, \rho)$ more generally, we see that the essential trends are well captured in the FLASH results. The alpha-off burn metrics are in excellent agreement in both 1D and 2D; larger differences are to be expected in the density where an inference was required for the HYDRA metrics. The most prominent discrepancy is the alpha-on temperature and pressure, where FLASH is not showing a significant increase in the burn-averaged pressure and the burn-averaged temperatures are likewise lower in FLASH. We are continuing to investigate these discrepancies; given the good agreement in the alpha-off metrics, the most likely explanation is differences in the alpha transport and energy deposition packages. FLASH's simple alpha energy diffusion model is likely to be substantially lower fidelity than the model used in the HYDRA calculations; several avenues exist for improving the alpha energy diffusion model in FLASH. Such improvements would affect the propagation of the fusion burn from the hot spot into the higher density ``shelf” and ultimately the degree of self-heating and alpha amplification of the fusion processes. Improvements will be made to our alpha energy models in the near future; nevertheless, we believe the level of agreement obtained with HYDRA is adequate for placing confidence in FLASH simulations of ignition scale targets.

Stepping back, this level of agreement between multiphysics codes for such highly integrated systems is noteworthy. The codes use different numerical methods and make many different modeling choices. It is deeply encouraging that FLASH is in such close agreement, and we believe this benchmark is compelling evidence for FLASH being considered validated for high performance magnetically driven ICF targets.

\section{Conclusion}

In aggregate, these six validation benchmarks indicate that, when properly configured, FLASH is capable of modeling magnetically driven ICF targets to a sufficient degree of accuracy to enable simulation-based target design. We have shown that FLASH accurately captures linear growth rates of key hydrodynamic instabilities in magnetically driven metallic liners, including magneto-Rayleigh-Taylor and Richtmeyer-Meshkov (Benchmarks 1, 2, 3), at the $\approx$20 MA scale of today's largest pulsed-power facility. Good agreement is also found with measurements of liquid deuterium compressed by a magnetically driven beryllium liner (Benchmark 4). With the addition of hydrogenic fusion reactions to Pacific Fusion's version of FLASH, we have shown good accuracy for modeling fusion burn metrics, including yield and burn-weighted ion temperatures, on one of the higher performing MagLIF experiments fielded at Z (Benchmark 5). Finally, although no experimental data is available at higher currents than accessible by Z, we have shown that FLASH agrees well with HYDRA simulations of similarity scaled MagLIF targets up to 60 MA in the absence of alpha energy feedback and to within factors of 2-3 in nuclear yield with alpha deposition on, showing ignition thresholds at similar levels of current as those predicted by HYDRA. 

This validated simulation capability will enable target design efforts at Pacific Fusion as we plan experiments for our upcoming Demonstration System, capable of delivering 60 MA to an appropriately impedance-matched load. Although code development remains active, both to incorporate new or improved physical models for enhanced accuracy and to accelerate time-to-solution, these validation benchmarks lend confidence that an essential set of physical models have been properly implemented and are capturing the dominant physical processes in these classes of targets. Because the Demonstration System is only a factor of three larger in current and roughly factor of ten larger in energy relative to Z, the dominant physical processes are expected to be quite similar across these machines. Of course, burning plasma dynamics including significant alpha heating and ignition will be novel to the Demonstration System; here, FLASH comparisons with HYDRA in the MagLIF scaling study lend confidence in light of HYDRA's successful modeling of igniting laser-driven ICF targets at the NIF \cite{Marinak_2024, Kritcher_2024}. 

\section{Acknowledgements}

The authors are grateful to Daniel Ruiz and Irvin Lindemuth for helpful discussions related to these benchmarks. 

The Flash Center for Computational Science acknowledges support by the U.S. Department of Energy National Nuclear Security Administration under Award Number DE-NA0004147, under subcontracts no. 630138 and C4574 with Los Alamos
National Laboratory. We also acknowledge support from the U.S. Department of Energy Advanced Research Projects Agency-Energy under Award Number DE-AR0001272 and the U.S. Department of Energy Office of Science under Award Number DE-SC0023246. The FLASH code used in this work was developed in part by the U.S. DOE NNSA- and the U.S. DOE Office of Science-supported Flash Center for Computational Science at the University of Chicago and the University of Rochester.

\bibliographystyle{unsrt}
\bibliography{refs}

\begin{thebibliography}{10}

\bibitem{Marinak_2024}
M.~M. Marinak, G.~B. Zimmerman, T.~Chapman, G.~D. Kerbel, M.~V. Patel, J.~M. Koning, S.~M. Sepke, B.~Chang, C.~R. Schroeder, J.~A. Harte, D.~S. Bailey, L.~A. Taylor, S.~H. Langer, M.~A. Belyaev, D.~S. Clark, J.~Gaffney, B.~A. Hammel, D.~E. Hinkel, A.~L. Kritcher, J.~L. Milovich, H.~F. Robey, and C.~R. Weber.
\newblock How numerical simulations helped to achieve breakeven on the {NIF}.
\newblock {\em Physics of Plasmas}, 31(7):070501, 07 2024.

\bibitem{Kritcher_2024}
A.~L. Kritcher, A.~B. Zylstra, C.~R. Weber, O.~A. Hurricane, D.~A. Callahan, D.~S. Clark, L.~Divol, D.~E. Hinkel, K.~Humbird, O.~Jones, J.~D. Lindl, S.~Maclaren, D.~J. Strozzi, C.~V. Young, A.~Allen, B.~Bachmann, K.~L. Baker, T.~Braun, G.~Brunton, D.~T. Casey, T.~Chapman, C.~Choate, E.~Dewald, J.-M.~G. Di~Nicola, M.~J. Edwards, S.~Haan, T.~Fehrenbach, M.~Hohenberger, E.~Kur, B.~Kustowski, C.~Kong, O.~L. Landen, D.~Larson, B.~J. MacGowan, M.~Marinak, M.~Millot, A.~Nikroo, R.~Nora, A.~Pak, P.~K. Patel, J.~E. Ralph, M.~Ratledge, M.~S. Rubery, D.~J. Schlossberg, S.~M. Sepke, M.~Stadermann, T.~I. Suratwala, R.~Tommasini, R.~Town, B.~Woodworth, B.~Van~Wonterghem, and C.~Wild.
\newblock {Design of the first fusion experiment to achieve target energy gain $G>1$}.
\newblock {\em Phys. Rev. E}, 109:025204, Feb 2024.

\bibitem{Ruiz_2023_Iscaling}
D.~E. Ruiz, P.~F. Schmit, D.~A. Yager-Elorriaga, M.~R. Gomez, M.~R. Weis, C.~A. Jennings, A.~J. Harvey-Thompson, P.~F. Knapp, S.~A. Slutz, D.~J. Ampleford, K.~Beckwith, and M.~K. Matzen.
\newblock Exploring the parameter space of {MagLIF} implosions using similarity scaling. {II}. {C}urrent scaling.
\newblock {\em Phys. Plasmas}, 30(3):032708, 2023.

\bibitem{Knapp_2017}
P.~F. Knapp, M.~R. Martin, D.~H. Dolan, K.~Cochrane, D.~Dalton, J.-P. Davis, C.~A. Jennings, G.~P. Loisel, D.~H. Romero, I.~C. Smith, E.~P. Yu, M.~R. Weis, T.~R. Mattsson, R.~D. McBride, K.~Peterson, J.~Schwarz, and D.~B. Sinars.
\newblock {Direct measurement of the inertial confinement time in a magnetically driven implosion}.
\newblock {\em Phys. Plasmas}, 24(4):042708, 04 2017.

\bibitem{Sinars_2011}
D.~B. Sinars, S.~A. Slutz, M.~C. Herrmann, R.~D. McBride, M.~E. Cuneo, C.~A. Jennings, J.~P. Chittenden, A.~L. Velikovich, K.~J. Peterson, R.~A. Vesey, C.~Nakhleh, E.~M. Waisman, B.~E. Blue, K.~Killebrew, D.~Schroen, K.~Tomlinson, A.~D. Edens, M.~R. Lopez, I.~C. Smith, J.~Shores, V.~Bigman, G.~R. Bennett, B.~W. Atherton, M.~Savage, W.~A. Stygar, G.~T. Leifeste, and J.~L. Porter.
\newblock {Measurements of magneto-Rayleigh–Taylor instability growth during the implosion of initially solid metal liners}.
\newblock {\em Phys. Plasmas}, 18(5):056301, 04 2011.

\bibitem{Ellison_2024}
C.L. Ellison, V.~Garcia, M.~Gomez, G.~P. Grim, J.~H. Hammer, C.~A. Jennings, P.~F. Knapp, K.~R. LeChien, N.~B. Meezan, R.~Peterson, A.~Reyes, A.~Steiner, W.~A. Stygar, P.~Tzeferacos, D.~Welch, and A.~B. Zylstra.
\newblock Opportunities in pulsed magnetic fusion energy, 2024.

\bibitem{AMPS_2025}
A.~Alexander, L.~R. Benedetti, I.~Bhattacharyya, J.~Bowen, J.~C. Cabatu, V.~Cacdac, F.~Chhavi, C.~Chen, K.~Chen, D.~Clark, J.~Clark, T.~Cope, W.~Dannemann, S.~Davidson, D.~DeHaan, J.~Dugan, M.~Eihusen, C.~L. Ellison, C.~Esquivel, D.~Ethridge, B.~Ferguson, B.~Ferguson, J.~Fry, F.~García-Rubio, T.~Goyal, G.~Grim, J.~Grodman, B.~Haid, F.~Howland, T.~Van~Huynh, V.~John, P.~F. Knapp, I.~Kravitz, E.~S. Lander, S.~Langendorf, K.~R. LeChien, A.~Link, N.~B. Meezan, D.~S. Miller, N.~Nardelli, Q.~Ogirri, J.~Peng, A.~Pinto, R.~Powser, F.~Puno, K.~Quang, B.~Rahn, W.~Regan, K.~Reichenbach, A.~Reyes, C.~Richardson, D.~Rose, J.~Samaniego, P.~F. Schmit, V.~Silva, N.~Simon, S.~Sitaraman, H.~Sullan, J.~Trebesch, M.~Truong, C.~von Muench, C.~Waltz, D.~Williams, E.~Wood, S.~Wu, and A.~B. Zylstra.
\newblock {Affordable, manageable, practical, and scalable (AMPS)} high-yield and high-gain inertial fusion.
\newblock {\em Submitted to Physics of Plasmas, Special Issue on Pulsed Magnetic Fusion Energy}, 2025.

\bibitem{Fryxell_2000}
B.~Fryxell, K.~Olson, P.~Ricker, F.~X. Timmes, M.~Zingale, D.~Q. Lamb, P.~MacNeice, R.~Rosner, J.~W. Truran, and H.~Tufo.
\newblock {FLASH}: An adaptive mesh hydrodynamics code for modeling astrophysical thermonuclear flashes.
\newblock {\em Astrophys. J. Suppl. Ser.}, 131(1):273--334, Nov 2000.

\bibitem{Tzeferacos_2015}
P.~Tzeferacos, M.~Fatenejad, N.~Flocke, C.~Graziani, G.~Gregori, D.~Q. Lamb, D.~Lee, J.~Meinecke, A.~Scopatz, and K.~Weide.
\newblock {FLASH MHD simulations of experiments that study shock-generated magnetic fields}.
\newblock {\em High Energy Density Phys.}, 17, Part A:24--31, 2015.

\bibitem{Tzeferacos_2018}
P.~Tzeferacos, A.~Rigby, A.~F.~A. Bott, A.~R. Bell, R.~Bingham, A.~Casner, F.~Cattaneo, E.~M. Churazov, J.~Emig, F.~Fiuza, et~al.
\newblock Laboratory evidence of dynamo amplification of magnetic fields in a turbulent plasma.
\newblock {\em Nat. Commun.}, 9(1):1--8, 2018.

\bibitem{Bott_2021}
Archie F.~A. Bott, Petros Tzeferacos, Laura Chen, Charlotte A.~J. Palmer, Alexandra Rigby, Anthony~R. Bell, Robert Bingham, Andrew Birkel, Carlo Graziani, Dustin~H. Froula, et~al.
\newblock Time-resolved turbulent dynamo in a laser plasma.
\newblock {\em Proc. Natl. Acad. Sci. U. S. A.}, 118(11), 2021.

\bibitem{Townsley_2007}
D.~M. Townsley, A.~C. Calder, S.~M. Asida, I.~R. Seitenzahl, F.~Peng, N.~Vladimirova, D.~Q. Lamb, and J.~W. Truran.
\newblock Flame evolution during type ia supernovae and the deflagration phase in the gravitationally confined detonation scenario.
\newblock {\em The Astrophysical Journal}, 668(2):1118, oct 2007.

\bibitem{Fatenejad_2013}
Milad Fatenejad, B.~Fryxell, J.~Wohlbier, E.~Myra, D.~Lamb, C.~Fryer, and C.~Graziani.
\newblock Collaborative comparison of simulation codes for high-energy-density physics applications.
\newblock {\em High Energy Density Phys.}, 9(1):63--66, 2013.

\bibitem{Sauppe_2023}
J.~P. Sauppe, Y.~Lu, P.~Tzeferacos, A.~C. Reyes, S.~Palaniyappan, K.~A. Flippo, S.~Li, and J.~L. Kline.
\newblock {On the importance of three-dimensional modeling for high-energy-density physics experiments}.
\newblock {\em Phys. Plasmas}, 30(6):062707, 06 2023.

\bibitem{Hansen_2024}
E.~C. Hansen, F.~Garcia-Rubio, M.~B.~P. Adams, M.~Fatenejad, K.~Moczulski, P.~Ney, H.~U. Rahman, A.~C. Reyes, E.~Ruskov, V.~Tranchant, and P.~Tzeferacos.
\newblock Feasibility and performance of the staged {Z}-pinch: A one-dimensional study with {FLASH} and {MACH2}.
\newblock {\em Physics of Plasmas}, 31(4):042712, 04 2024.

\bibitem{Michta_2024}
D.~Michta, V.~Tranchant, J.~Narkis, E.~C. Hansen, M.~B.~P. Adams, K.~Moczulski, A.~Reyes, F.~Conti, F.~N. Beg, and P.~Tzeferacos.
\newblock Code-to-code comparison between {FLASH} and {HYDRA} in gas-puff {Z}-pinch modeling.
\newblock {\em Physics of Plasmas}, 31(12):122703, 12 2024.

\bibitem{Tranchant_2025}
V.~Tranchant, E.~C. Hansen, D.~Michta, F.~Garcia-Rubio, H.~U. Rahman, P.~Ney, E.~Ruskov, and P.~Tzeferacos.
\newblock A two-dimensional numerical study of the magneto-{R}ayleigh–{T}aylor instability with {FLASH}: {A}pplication to the staged {Z}-pinch concept.
\newblock {\em Physics of Plasmas}, 32(3):033901, 03 2025.

\bibitem{Lee_2013}
Dongwook {Lee}.
\newblock A solution accurate, efficient and stable unsplit staggered mesh scheme for three dimensional magnetohydrodynamics.
\newblock {\em J. Computat. Phys.}, 243:269--292, jun 2013.

\bibitem{Tzeferacos_2012}
P.~Tzeferacos, M.~Fatenejad, N.~Flocke, G.~Gregori, D.Q. Lamb, D.~Lee, J.~Meinecke, A.~Scopatz, and K.~Weide.
\newblock {FLASH} magnetohydrodynamic simulations of shock-generated magnetic field experiments.
\newblock {\em High Energy Density Physics}, 8(4):322--328, 2012.

\bibitem{Ji_2013}
Jeong-Young Ji and Eric~D. Held.
\newblock Closure and transport theory for high-collisionality electron-ion plasmas.
\newblock {\em Phys. Plasmas}, 20(4):042114, 2013.

\bibitem{Tzeferacos_2017}
P~Tzeferacos, A.~Rigby, A.~Bott, A.~R. Bell, Robert Bingham, A.~Casner, F.~Cattaneo, E.~M. Churazov, J.~Emig, N.~Flocke, et~al.
\newblock Numerical modeling of laser-driven experiments aiming to demonstrate magnetic field amplification via turbulent dynamo.
\newblock {\em Phys. Plasmas}, 24(4):041404, 2017.

\bibitem{FLASH_UserGuide_2024}
Flash~Center at~the University~of Rochester.
\newblock {\em FLASH User Guide}, 2024.

\bibitem{Sinars_2010}
D.~B. Sinars, S.~A. Slutz, M.~C. Herrmann, R.~D. McBride, M.~E. Cuneo, K.~J. Peterson, R.~A. Vesey, C.~Nakhleh, B.~E. Blue, K.~Killebrew, D.~Schroen, K.~Tomlinson, A.~D. Edens, M.~R. Lopez, I.~C. Smith, J.~Shores, V.~Bigman, G.~R. Bennett, B.~W. Atherton, M.~Savage, W.~A. Stygar, G.~T. Leifeste, and J.~L. Porter.
\newblock {Measurements of Magneto-Rayleigh-Taylor Instability Growth during the Implosion of Initially Solid Al Tubes Driven by the 20-MA, 100-ns Z Facility}.
\newblock {\em Phys. Rev. Lett.}, 105:185001, Oct 2010.

\bibitem{McBride_2012}
R.~D. McBride, S.~A. Slutz, C.~A. Jennings, D.~B. Sinars, M.~E. Cuneo, M.~C. Herrmann, R.~W. Lemke, M.~R. Martin, R.~A. Vesey, K.~J. Peterson, A.~B. Sefkow, C.~Nakhleh, B.~E. Blue, K.~Killebrew, D.~Schroen, T.~J. Rogers, A.~Laspe, M.~R. Lopez, I.~C. Smith, B.~W. Atherton, M.~Savage, W.~A. Stygar, and J.~L. Porter.
\newblock Penetrating radiography of imploding and stagnating beryllium liners on the $z$ accelerator.
\newblock {\em Phys. Rev. Lett.}, 109:135004, Sep 2012.

\bibitem{McBride_2013}
R.~D. McBride, M.~R. Martin, R.~W. Lemke, J.~B. Greenly, C.~A. Jennings, D.~C. Rovang, D.~B. Sinars, M.~E. Cuneo, M.~C. Herrmann, S.~A. Slutz, C.~W. Nakhleh, D.~D. Ryutov, J.-P. Davis, D.~G. Flicker, B.~E. Blue, K.~Tomlinson, D.~Schroen, R.~M. Stamm, G.~E. Smith, J.~K. Moore, T.~J. Rogers, G.~K. Robertson, R.~J. Kamm, I.~C. Smith, M.~Savage, W.~A. Stygar, G.~A. Rochau, M.~Jones, M.~R. Lopez, J.~L. Porter, and M.~K. Matzen.
\newblock Beryllium liner implosion experiments on the z accelerator in preparation for magnetized liner inertial fusion.
\newblock {\em Physics of Plasmas}, 20(5):056309, 05 2013.

\bibitem{Knapp_2020}
P.~F. Knapp, M.~R. Martin, D.~Yager-Elorriaga, A.~J. Porwitzky, F.~W. Doss, G.~A. Shipley, C.~A. Jennings, D.~E. Ruiz, T.~Byvank, C.~C. Kuranz, C.~E. Myers, D.~H. Dolan, K.~Cochrane, M.~Schollmeier, I.~C. Smith, T.~R. Mattsson, B.~M. Jones, K.~Peterson, J.~Schwarz, R.~D. McBride, D.~G. Flicker, and D.~B. Sinars.
\newblock {A novel, magnetically driven convergent Richtmyer–Meshkov platform}.
\newblock {\em Phys. Plasmas}, 27(9):092707, 09 2020.

\bibitem{Lewis_2023}
William~E. Lewis, Owen~M. Mannion, D.~E. Ruiz, Christopher~A. Jennings, Patrick~F. Knapp, Matthew~R. Gomez, Adam~J. Harvey-Thompson, Matthew~R. Weis, Stephen~A. Slutz, David~J. Ampleford, and Kristian Beckwith.
\newblock Data-driven assessment of magnetic charged particle confinement parameter scaling in magnetized liner inertial fusion experiments on {Z}.
\newblock {\em Physics of Plasmas}, 30(5):052701, 05 2023.

\bibitem{Lee_1984}
Yim~T Lee and RM~More.
\newblock An electron conductivity model for dense plasmas.
\newblock {\em The Physics of Fluids}, 27(5):1273--1286, 1984.

\bibitem{slutz2010pulsed}
SA~Slutz, MC~Herrmann, RA~Vesey, AB~Sefkow, DB~Sinars, DC~Rovang, KJ~Peterson, and ME~Cuneo.
\newblock Pulsed-power-driven cylindrical liner implosions of laser preheated fuel magnetized with an axial field.
\newblock {\em Physics of Plasmas}, 17(5), 2010.

\bibitem{Magee_1995}
N.~Magee, J.~Jr. Abdallah, R.~Clark, J.~Cohen, L.~Collins, George Csanak, C.~Fontes, A.~Gauger, J.~Keady, D.~Kilcrease, and A.~Merts.
\newblock Atomic structure calculations and new {Los Alamos} astrophysical opacities.
\newblock {\em Astrophysical Applications of Powerful New Databases}, 78:51, 01 1995.

\bibitem{BoschHale1992}
H.-S. Bosch and G.M. Hale.
\newblock Improved formulas for fusion cross-sections and thermal reactivities.
\newblock {\em Nuclear Fusion}, 32(4):611, apr 1992.

\bibitem{Falgout_2006}
Robert~D. Falgout, Jim~E. Jones, and Ulrike~Meier Yang.
\newblock The design and implementation of hypre, a library of parallel high performance preconditioners.
\newblock In Are~Magnus Bruaset and Aslak Tveito, editors, {\em Numerical Solution of Partial Differential Equations on Parallel Computers}, pages 267--294, Berlin, Heidelberg, 2006. Springer Berlin Heidelberg.

\bibitem{Slutz_PoP_2010}
S.~A. Slutz, M.~C. Herrmann, R.~A. Vesey, A.~B. Sefkow, D.~B. Sinars, D.~C. Rovang, K.~J. Peterson, and M.~E. Cuneo.
\newblock Pulsed-power-driven cylindrical liner implosions of laser preheated fuel magnetized with an axial field.
\newblock {\em Phys. Plasmas}, 17(5):056303, 2010.

\bibitem{Braginskii1965}
S.~I. Braginskii.
\newblock Transport processes in a plasma.
\newblock {\em Rev. Plasma Phys.}, 1:205, 1965.

\bibitem{Olson_1999}
KM~Olson, P~MacNeice, B~Fryxell, P~Ricker, FX~Timmes, and M~Zingale.
\newblock {PARAMESH}: a parallel, adaptive mesh refinement toolkit and performance of the {ASCI}/{FLASH} code.
\newblock In {\em American Astronomical Society Meeting Abstracts}, volume 195, pages 42--03, 1999.

\bibitem{MacNeice_2000}
Peter MacNeice, Kevin~M. Olson, Clark Mobarry, Rosalinda {de Fainchtein}, and Charles Packer.
\newblock Paramesh: A parallel adaptive mesh refinement community toolkit.
\newblock {\em Computer Physics Communications}, 126(3):330--354, 2000.

\bibitem{BergerOliger1984}
Marsha~J Berger and Joseph Oliger.
\newblock Adaptive mesh refinement for hyperbolic partial differential equations.
\newblock {\em Journal of Computational Physics}, 53(3):484--512, 1984.

\bibitem{BergerColella1989}
M.J. Berger and P.~Colella.
\newblock Local adaptive mesh refinement for shock hydrodynamics.
\newblock {\em Journal of Computational Physics}, 82(1):64--84, 1989.

\bibitem{Khokhlov1998}
A.M Khokhlov.
\newblock Fully threaded tree algorithms for adaptive refinement fluid dynamics simulations.
\newblock {\em Journal of Computational Physics}, 143(2):519--543, 1998.

\bibitem{Lohner_1987}
Rainald L{\"o}hner.
\newblock An adaptive finite element scheme for transient problems in {CFD}.
\newblock {\em Computer methods in applied mechanics and engineering}, 61(3):323--338, 1987.

\bibitem{Godunov_1959}
Sergei~K. Godunov and I.~Bohachevsky.
\newblock {Finite difference method for numerical computation of discontinuous solutions of the equations of fluid dynamics}.
\newblock {\em {Matemati{\v c}eskij sbornik}}, 47(89)(3):271--306, 1959.

\bibitem{Colella_1984}
Phillip Colella and Paul~R Woodward.
\newblock The {P}iecewise {P}arabolic {M}ethod ({PPM}) for gas-dynamical simulations.
\newblock {\em Journal of Computational Physics}, 54(1):174--201, 1984.

\bibitem{Colella_1990}
Phillip Colella.
\newblock Multidimensional upwind methods for hyperbolic conservation laws.
\newblock {\em Journal of Computational Physics}, 87(1):171--200, 1990.

\bibitem{Lee_2009}
Dongwook Lee and Anil~E. Deane.
\newblock An unsplit staggered mesh scheme for multidimensional magnetohydrodynamics.
\newblock {\em Journal of Computational Physics}, 228(4):952--975, 2009.

\bibitem{Evans_1988}
Charles~R Evans and John~F Hawley.
\newblock Simulation of magnetohydrodynamic flows-a constrained transport method.
\newblock {\em Astrophysical Journal, Part 1 (ISSN 0004-637X), vol. 332, Sept. 15, 1988, p. 659-677.}, 332:659--677, 1988.

\bibitem{Kruskal_1954}
Martin~David Kruskal, Martin Schwarzschild, and Subrahmanyan Chandrasekhar.
\newblock Some instabilities of a completely ionized plasma.
\newblock {\em Proceedings of the Royal Society of London. Series A. Mathematical and Physical Sciences}, 223(1154):348--360, 1954.

\bibitem{Harris_1962}
E.~G. Harris.
\newblock Rayleigh‐{T}aylor instabilities of a collapsing cylindrical shell in a magnetic field.
\newblock {\em The Physics of Fluids}, 5(9):1057--1062, 09 1962.

\bibitem{Sinars_2012}
D.~B. Sinars, R.~D. McBride, D.~Rovang, A.~Sefkow, S.~Slutz, R.~Lemke, M.~Cuneo, M.~Herrmann, C.~Jennings, M.~Jobe, D.~Lamppa, M.~Martin, C.~Nakhleh, A.~Owen, J.~McKenney, R.~Mock, T.~Peters, G.~Torres, and E.~Waisman.
\newblock Stability of fusion target concepts on {Z}. {SAND2012-8009}.
\newblock Technical report, Sandia National Laboratories, 2012.

\bibitem{Yu_2023}
E.~P. Yu, T.~J. Awe, K.~R. Cochrane, K.~J. Peterson, K.~C. Yates, T.~M. Hutchinson, M.~W. Hatch, B.~S. Bauer, K.~Tomlinson, and D.~B. Sinars.
\newblock Seeding the electrothermal instability through a three-dimensional, nonlinear perturbation.
\newblock {\em Phys. Rev. Lett.}, 130:255101, Jun 2023.

\bibitem{Peterson_2013}
K.~J. Peterson, Edmund~P. Yu, Daniel~B. Sinars, Michael~E. Cuneo, Stephen~A. Slutz, Joseph~M. Koning, Michael~M. Marinak, Charles Nakhleh, and Mark~C. Herrmann.
\newblock {Simulations of electrothermal instability growth in solid aluminum rods}.
\newblock {\em Phys. Plasmas}, 20(5):056305, 04 2013.

\bibitem{Hutchinson_2020}
Trevor~M. Hutchinson.
\newblock {\em Experimental studies of the electrothermal instability using electrically thick, coated metal}.
\newblock PhD thesis, University of Nevada, Reno, 2020.

\bibitem{Knapp_2019}
P.~F. Knapp, M.~R. Gomez, S.~B. Hansen, M.~E. Glinsky, C.~A. Jennings, S.~A. Slutz, E.~C. Harding, K.~D. Hahn, M.~R. Weis, M.~Evans, M.~R. Martin, A.~J. Harvey-Thompson, M.~Geissel, I.~C. Smith, D.~E. Ruiz, K.~J. Peterson, B.~M. Jones, J.~Schwarz, G.~A. Rochau, D.~B. Sinars, R.~D. McBride, and P.-A. Gourdain.
\newblock Origins and effects of mix on magnetized liner inertial fusion target performance.
\newblock {\em Physics of Plasmas}, 26(1):012704, 01 2019.

\bibitem{Gomez_2020}
M.~R. Gomez, S.~A. Slutz, C.~A. Jennings, D.~J. Ampleford, M.~R. Weis, C.~E. Myers, D.~A. Yager-Elorriaga, K.~D. Hahn, S.~B. Hansen, E.~C. Harding, A.~J. Harvey-Thompson, D.~C. Lamppa, M.~Mangan, P.~F. Knapp, T.~J. Awe, G.~A. Chandler, G.~W. Cooper, J.~R. Fein, M.~Geissel, M.~E. Glinsky, W.~E. Lewis, C.~L. Ruiz, D.~E. Ruiz, M.~E. Savage, P.~F. Schmit, I.~C. Smith, J.~D. Styron, J.~L. Porter, B.~Jones, T.~R. Mattsson, K.~J. Peterson, G.~A. Rochau, and D.~B. Sinars.
\newblock Performance scaling in magnetized liner inertial fusion experiments.
\newblock {\em Phys. Rev. Lett.}, 125:155002, Oct 2020.

\bibitem{Lewis_2021}
William~E. Lewis, Patrick~F. Knapp, Stephen~A. Slutz, Paul~F. Schmit, Gordon~A. Chandler, Matthew~R. Gomez, Adam~J. Harvey-Thompson, Michael~A. Mangan, David~J. Ampleford, and Kristian Beckwith.
\newblock {Deep-learning-enabled Bayesian inference of fuel magnetization in magnetized liner inertial fusion}.
\newblock {\em Physics of Plasmas}, 28(9):092701, 09 2021.

\bibitem{Ampleford_2024}
D.~J. Ampleford, D.~A. Yager-Elorriaga, C.~A. Jennings, E.~C. Harding, M.~R. Gomez, A.~J. Harvey-Thompson, T.~J. Awe, G.~A. Chandler, G.~S. Dunham, M.~Geissel, K.~D. Hahn, S.~B. Hansen, P.~F. Knapp, D.~C. Lamppa, W.~E. Lewis, L.~Lucero, M.~Mangan, R.~Paguio, L.~Perea, G.~A. Robertson, C.~L. Ruiz, D.~E. Ruiz, P.~F. Schmit, S.~A. Slutz, G.~E. Smith, I.~C. Smith, C.~S. Speas, T.~J. Webb, M.~R. Weis, K.~Whittemore, E.~P. Yu, R.~D. McBride, K.~J. Peterson, B.~M. Jones, G.~A. Rochau, and D.~B. Sinars.
\newblock Controlling morphology and improving reproducibility of magnetized liner inertial fusion experiments.
\newblock {\em Physics of Plasmas}, 31(2):022703, 02 2024.

\bibitem{Hutchinson_2023}
T.~M. Hutchinson, S.~J. Ali, G.~P. Grim, N.~B. Meezan, and K.~R. LeChien.
\newblock Bayesian inferences of electrical current delivered to shocked transmission lines.
\newblock {\em Journal of Applied Physics}, 134(15):155901, 10 2023.

\bibitem{Lewis_2021_ICOPS}
William~E. Lewis, Patrick~F. Knapp, Stephen~A. Slutz, Paul~F. Schmit, Gordon~A. Chandler, Matthew~R. Gomez, Adam~J. Harvey-Thompson, Michael~A. Mangan, David~J. Ampleford, and Kristian Beckwith.
\newblock Deep learning enabled assessment of magnetic confinement in magnetized liner inertial fusion.
\newblock In {\em 2021 IEEE International Conference on Plasma Science (ICOPS)}, pages 1--1, 2021.

\end{thebibliography}

\end{document}